\documentclass[manuscript]{aastex}

\pdfoutput=1

\slugcomment{Not to appear in Nonlearned J., 45.}

\shorttitle{A Catalog of Visually Classified Galaxies}
\shortauthors{Ann , Seo \& Ha}

\begin{document}


\title{A Catalog of Visually Classified Galaxies in the Local ($z\sim0.01$)
Universe}


\author{H. B. Ann\altaffilmark{1} and Mira Seo\altaffilmark{1}}
\affil{Department of Earth Science, Pusan National University,
    Busan, Korea}

\and

\author{D. K. Ha\altaffilmark{2}}
\affil{Youngdo High School, Busan, Korea}
\email{hbann@pusan.ac.kr}

\begin{abstract}
The morphological types of 5836 galaxies were classified by a visual 
inspection of color images using the Sloan Digital Sky
Survey (SDSS) Data Release 7 (DR7) to produce a morphology catalog of 
a representative sample of local galaxies with $z<0.01$. 
The sample galaxies are almost complete for galaxies brighter than 
$r_{pet}=17.77$. Our classification system is basically the same as that of
the Third Reference Catalog of Bright Galaxies with some simplifications for
giant galaxies. On the other hand, we distinguish the fine features of dwarf
elliptical (dE)-like galaxies to classify five subtypes: dE, blue-cored dwarf
ellipticals, dwarf spheroidals (dSph), blue dwarf 
ellipticals (dE$_{blue}$), and dwarf lenticulars (dS0). In addition, we denote
the presence of nucleation in dE, dSph, and dS0. 
Elliptical galaxies and lenticular galaxies contribute only $\sim1.5\%$
and $\sim4.9\%$ of the local galaxies, respectively, whereas spirals and
irregulars contribute $\sim32.1\%$ and $\sim42.8\%$, respectively. The 
dE$_{blue}$ galaxies, which are a recently discovered population of galaxies,  
contribute a significant fraction of the dwarf galaxies. There seem to be 
structural differences between dSph and dE galaxies. The dSph galaxies are
fainter and bluer with a shallower surface brightness gradient than dE galaxies.
They also have a lower fraction of galaxies with small axis 
ratios ($b/a \lesssim 0.4$) than dE galaxies.
The mean projected distance to the nearest neighbor galaxy is $\sim260$kpc. 
About $1\%$ of local galaxies have no neighbors with comparable luminosity
within a projected distance of 2Mpc.

\end{abstract}



\section{Introduction}

The morphology of galaxies is of great interest because it reflects not only
the structural properties of galaxies but also their star formation histories.
A morphological study of galaxies has been one of the main topics of 
extra-galactic studies since the pioneering work by \citet{hub36}. 
Despite the enormous efforts to determine the morphological types of galaxies,
the morphological types of the majority of galaxies from recent surveys, such 
as the Sloan Digital Sky Survey (SDSS), still need to be determined.

The morphology of bright galaxies in the local universe has been well studied
and summarized in catalogs, such as A Revised Shaply-Ames Catalog of Bright 
Galaxies (RSA, \citet{san81}) and the Third Reference Catalog of Bright
Galaxies (RC3, \citet{deV91}). However, none of these catalogs are suitable for 
dwarf galaxy studies because most dwarf galaxies which are much 
fainter than the bright end of dwarf galaxies are omitted due to their 
selection criteria.
The recent study of \citet{nai10}, who visually classified 14,000 SDSS galaxies
from Data Release 4 (DR4), focussed on the redshift range $0.01 < z <0.10$,
and as a consequence included few dwarf galaxies. The Galaxy
Zoo Project \citep{lin08} has provided morphological
information for a much larger number of objects, but only in broad
categories with little information on dwarf galaxies.

Dwarf galaxies, however, are predicted to be the most dominant type of galaxy
based on cold dark matter (CDM) cosmology \citep{whi78}. The importance of
dwarf galaxies in galactic astronomy has been well demonstrated 
by the dominance of dwarf galaxies in the Local Group (LG), in which the number
of dwarf galaxies tripled recently, mostly thanks to SDSS \citep{mcc12,wal13}. 
They also dominate the nearby satellite systems such as those hosted
by M81 \citep{chi09} and M106 \citep{kim11}. Galaxies with distances less than 
$\sim11$Mpc and velocities relative to the LG ($V_{\rm{LG}}$) less than 
600kms$^{-1}$, which were compiled in the Updated Nearby Galaxy 
Catalog \citep{kar13a},
are also dominated by dwarf galaxies. \citet{mak11} presented an all-sky 
catalog of nearby galaxy groups that included 10,914 galaxies in the local 
universe ($z \lesssim0.01$) with various astrophysical parameters including the
morphological types in numerical code $\it{T}$. On the other hand, 
they did not report the detailed morphological types of dwarf galaxies.
\citet{kar13a} provided a better treatment of the morphology of
dwarf galaxies, but they confined the dwarfs to within $\sim11$Mpc.
Because galaxies in the local universe will provide
a unique sample to explore the structural properties of dwarf galaxies,
it is important to determine their detailed morphological types.

SDSS opens a new era of galaxy morphology by providing color images which are
useful for obtaining information about the underlying stellar populations. 
Because stellar populations are closely related to the star formation history 
of galaxies, the morphological types that consider the stellar populations
using color information and the light distributions can be considered to be
the best proxy for characterizing the integral property of galaxies. The blue 
elliptical galaxy \citep{str01},
 is an example of a new type of galaxy that has emerged
in the new era of the galaxy morphology. Moreover, it is deep enough to cover 
dwarf galaxies as faint as $M_{r}\approx-10$ in the local universe.

Another major difference between the morphological properties observed in
single-band images and color images were found in the dwarf elliptical (dE)
family. Although dE galaxies and dwarf spheroidal galaxies (dSph) are
distinguished in the LG, they were used interchangeably for
galaxies outside the LG in previous studies. In particular,
\citet{san84} used 'dE' and 'dE,N' for both dE galaxies and dSph
galaxies in a classification of Virgo Cluster
dwarfs, whereas \citet{kor12} used Sph for both galaxies. On the other hand, 
there appears to be some difference between dE and
dSph \citep{wei11,mcc12}. For example, dE
galaxies are generally brighter and redder than dSph galaxies. 

Although the environmental dependence of giant galaxies is well understood
thanks to recent surveys, such as SDSS \citep{got03,par07}, 
the dependence of dwarf galaxy morphology on environment
has not been well studied due to the lack of
dwarf galaxies in previous analyses. Therefore, it is important to produce a 
homogeneous sample of dwarf galaxies with detailed morphological types.
Because the limiting magnitude of spectroscopic observations of the SDSS is
$r=17.77$, it is possible to make a volume-limited sample brighter than 
$M_{r}=-15.2$ if galaxies with redshift less than $z=0.01$ are used. 
This provides an unprecedented opportunity to classify the
morphological types of local galaxies using the color images of the SDSS DR7
with an emphasis on dwarf galaxies, whose morphological properties are still
not well understood.
Particular attention was paid to the morphologies of dE-like
galaxies because they are the building blocks of larger structures in 
CDM cosmology. In particular, dSph are believed to be
mostly dark matter dominated dwarfs (e.g., Gilmore et al. 2007). 
Because dE galaxies show different properties from dSph,
such as the presence of gas and young stellar populations, 
at least for the LG dwarfs \citep{geh10}, it is important to determine whether
they can be distinguished by morphology alone. 

The aim of this paper is to perform a census of the morphology distribution 
of local galaxies brighter than $M_{r}\approx-15$.
To accomplish this, the morphological types of all SDSS galaxies that 
have redshifts less than $z=0.01$ are classified
by visual inspection of the color images of SDSS DR7. The results 
are presented in the form of a catalog that includes not only our 
classificiations, but also numerical T types,
adopted distances, absolute $r$-band magnitudes, $u-r$ colors, and
isophotal angular diameters and axis ratios. An additional aim is to
better understand the environment of local galaxies, a factor which can
play an important role in morphology.

Section 2 introduces the observational data used in this study and 
the procedure for selecting the sample galaxies. Section 3 describes the
methodology of morphology classification and Section 4 provides a 
catalog of the morphological types of the local galaxies. Section 5 presents
the physical and morphological properties of the local galaxies 
and Section 6 reports the 
environmental dependence of the morphology. The final section provides a 
brief summary and discussion. 

\section{Sample}

The basic sample of the local galaxies in this paper was derived from the
Korea Institute for Advanced Study Value-Added Galaxy 
Catalog (KIAS-VAGC Choi et al. 2010), but was supplemented by
galaxies with $z < 0.01$ in the NASA Extragalactic Data
Base (NED)\footnote{NED is operated by the Jet Propulsion Laboratory, 
California Institute of Technology, under contract with the National 
Aeronautics and Space Administration.} and the galaxies listed 
in \citet{mak11} that are not overlapped with the galaxies from the KIAS-VAGC
and NED. The KIAS-VAGC is a value-added catalog of galaxies based on the
New York University Value-added Catalog (NYU-VAGC) DR7 which was 
derived from the SDSS DR7 \citep{aba09}). the KIAS-VAGC 
provides valuable astrophysical parameters such as absolute magnitude
and colors corrected for the galactic extinction as well as distances
and morphological types determined by automated classifier \citep{par05}.
It also lists the redshifts of
10,497 galaxies that are not included in NYU-VAGC. They gathered redshift data
from various catalogs (see Choi et al. (2010) for a detailed description).
More detailed descriptions of SDSS data were reported by \citet{yor00} and
\citet{sto02}.

All of the galaxies with redshifts less than 0.01 were selected from the
KIAS-VAGC. The color images of the individual galaxies listed in the KIAS-VAGC
were examined carefully to determine whether they are parts of a galaxy or
multiply listed. A few tens of galaxies were removed from the selected galaxies
because of their multiple entry or invalid redshift. The
number of cleaned galaxies from KIAS-VAGC is 4807. In addition, 897 galaxies
from NED, whose images and photometric data
are available from SDSS DR7, were added. 
We present the number distribution of the sample galaxies as a function
of redshift in Figure 1.
The total number of sample galaxies is 5836.

%%%%%%%%%%%%%%%%%%%%%%%%%%%%%% Fig~1 (Hz.eps) %%%%%%%%%%%%%%%%%%%%%%
\begin{figure}
\epsscale{.80}
%\plotone{annfg1.eps}
\plotone{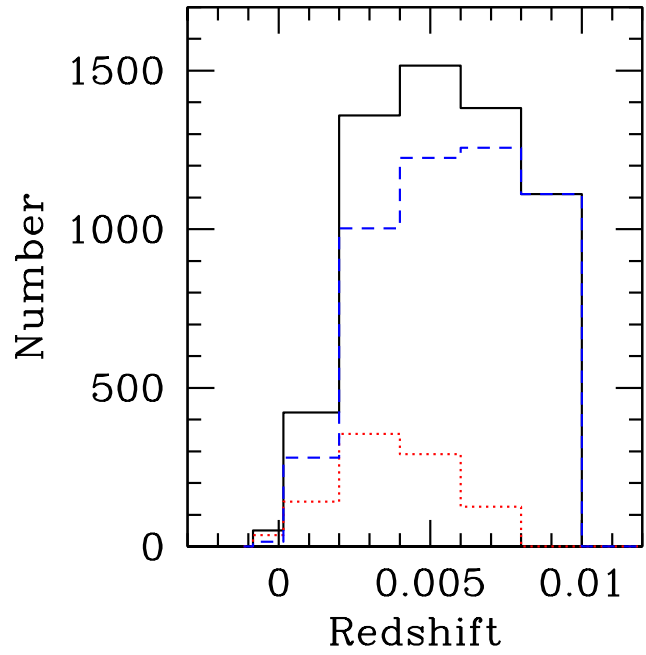}
\caption{Number distributions of local galaxies as a function of the
redshift. The dotted line indicates the galaxies in the Virgo cluster and the
short dashed line represents the remaining galaxies. The entire sample of
local galaxies is represented by the solid line.\label{fig1}
}
\end{figure}
%%%%%%%%%%%%%%%%%%%%%%%%%%%%%%%%%%%%%%%%%%%%%%%%%%%%%%%%%%%%%%%%%%%%%
\clearpage

Figure 2 shows the distribution of sample galaxies in a color-magnitude diagram.
The Petrosian $r$ magnitude and $u-r$ color are used as the representative
magnitude and color of a galaxy because the Petrosian $r$ magnitude was used
to select the spectroscopic target galaxies and the $u-r$ color is one of the
best proxies of the galaxy morphology because of its close dependence on the
star formation rate \citep{str01}.
Most galaxies brighter than the limiting magnitude of the SDSS spectroscopic
target galaxies ($r=17.77$) are located in a relatively narrow
range of $u-r$ colors, whereas galaxies fainter than $r=17.77$ have a wide range 
of $u-r$ colors. The color range of galaxies brighter than $r=17.77$ is similar
to that of the volume-limited sample of galaxies brighter 
than $M_{r}=-18.5$ \citep{cho07}. The redshifts of galaxies fainter than
$r=17.77$ have been observed by other
surveys, such as 2dFGRS \citep{col01}, which is deeper than SDSS.

%%%%%%%%%%%%%%%%%%%%%%%%%%%%%% Fig~2 (rur.ps  --> petrur.ps) %%%%%%%%%%%%%%%%%%%%%%%%%
\begin{figure}
%\tiny
%\centering
%\epsfxsize=7cm \epsfbox{rur.eps}
\epsscale{.80}
%\plotone{annfg2.eps}
\plotone{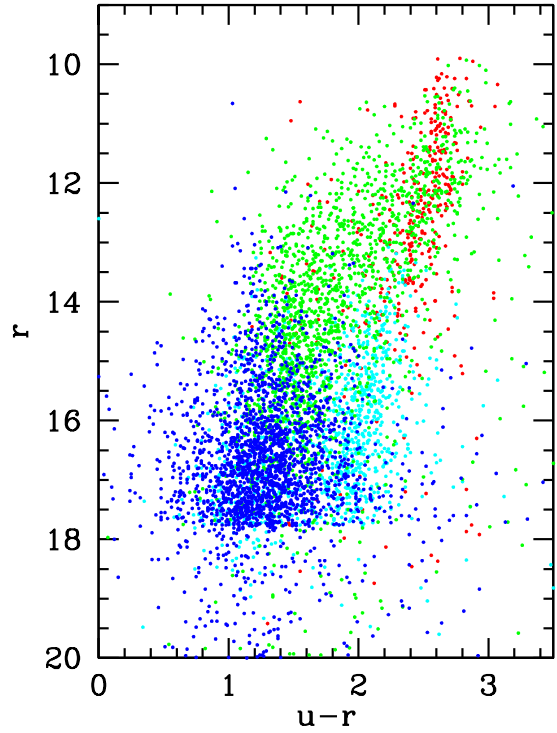}
\caption{Color-magnitude diagram of the sample galaxies. 
The apparent Petrosian magnitude and $u-r$ color are used.
We distinguish galaxies with different morphology 
by assigning different color codes: red (ellipticals and lenticulars), 
cyan (dEs), green (spirals), blue (irregulars).
\label{fig2}
}
\end{figure}
%\end{verbatim}
%%%%%%%%%%%%%%%%%%%%%%%%%%%%%%%%%%%%%%%%%%%%%%%%%%%%%%%%%%%%%%%%%%%%%
\clearpage

Figure 3 presents the distribution of $r$-band isophotal semi-major axis 
length ($a$) and the axial ratio ($b/a$) 
of sample galaxies whose isophotal semi-major and semi-minor
axis lengths are available in the SDSS DR7. The isophotal axis lengths were 
measured at an isophote of 25 mag arcsec$^{-2}$ at each pass-band.
Most galaxies
are larger than $a_{iso}\approx20$ pixels. Therefore, galaxies in the local
universe ($z < 0.01$) are large enough for visual classification. 
Approximately $\sim330$ spiral galaxies are inclined more 
than $\sim73^{\circ}$. 
These edge-on galaxies comprise $\sim5\%$ of the total galaxies and $\sim20\%$ 
of spiral galaxies. Galaxies less than $\sim20$arcsec are mostly
face-on galaxies.

%%%%%%%%%%%%%% Fig~3 (isoA_b2a.ps) %%%%%%%%%%%%%%%%%%%%
\begin{figure}
%\epsscale{.80}
%\plotone{annfg3.eps}
\plotone{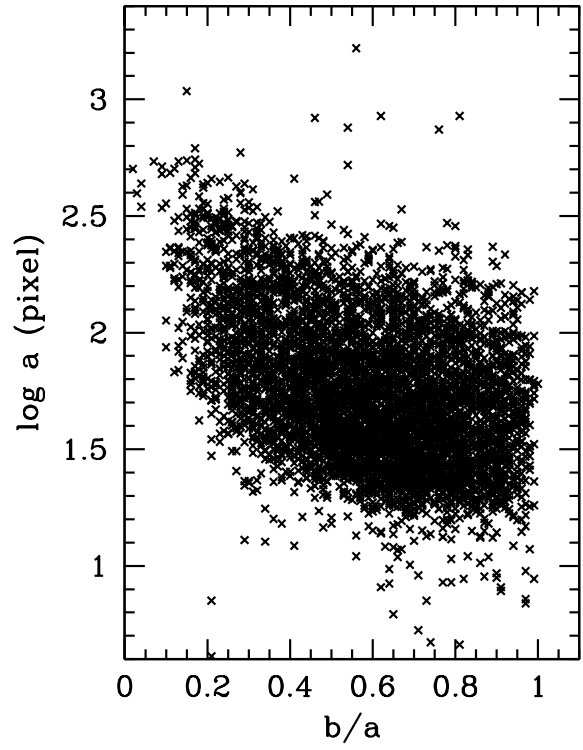}
\caption{Semi-major axis and axial ratio of the sample galaxies. The $r$-band 
isophotal semi-major and semi-minor axes are used to derive axial ratios.
\label{fig3}
}
\end{figure}
%%%%%%%%%%%%%%%%%%%%%%%%%%%%%%%%%%%%%%%%%%%%%%%%%%%%%%%%%%%%%%%%%%%%%
\clearpage

\section{Morphology Classification}

Traditionally, morphological classification has been performed through
inspection of photographic plates or prints sensitive to the blue
region of the spectrum (e.g.,  de Vaucouleurs 1959; Sandage 1961; 
Sandage and Tammann 1981; Sandage and Bedke 1994). More recently,
galaxies have been classified using logarithmic, sky-subtracted
digital $B$-band images in units of mag arcsec$^{-2}$ (e.g., Buta
et al. 2007). Therefore, in order to be "true" to the traditional
classification systems, all of which are built on the original
Hubble (1926, 1936) system, one should use SDSS $g$-band images
since these are close enough in wavelength to the $B$ band.

In the present study, however, we use the color images provided by
the SDSS which are based on a combination of $gri$ images \citep{lup04}.
For some galaxy types, the use of such images could
lead to systematic differences from blue light classifications
because the inclusion of the $i$ band can enhance the prominence
of galactic bulges, which play a role in the T type.
Our reasons for using color images over $g$-band images are twofold.
First, color adds information to morphology by its ability to
distinguish underlying stellar populations. Second, our sample
is dominated by galaxies that have very little or no bulge, making
the possible impact of the $i$ band on T types less of an issue.
\citet{par05} show that color gradients and the colors of galaxies
are effective for automatic classification. We therefore feel that the use
of color images for visual classification is justified.

%%%%%%%%%%%%%% Fig~4 (giant.eps  (EspIm.eps)   ) %%%%%%%%%%%%%%%%%%%%
\begin{figure}
\centering
%\includegraphics[scale=1.00]{giant.eps}
%\plotone{annfg4.eps}
\plotone{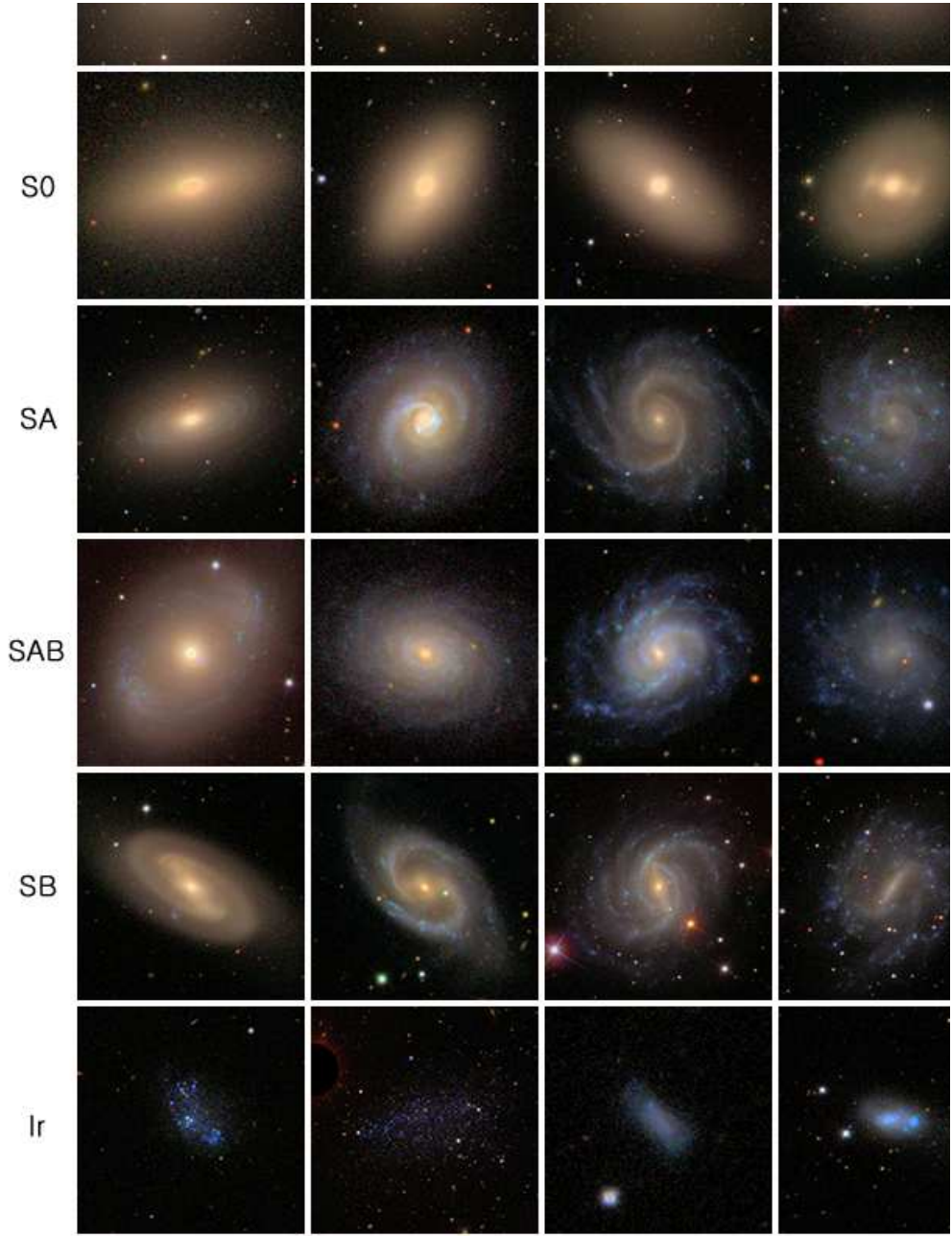}
\caption{Sample images of E, S0, Sp, and Ir galaxies. 
Elliptical and lenticulars are in the first and second rows, respectively.
Spirals grouped according to the bar types (SA, SAB, and SB) are displayed in
the third, fourth, and fifth rows, respectively.
Irregular galaxies, including BCDs are presented in the
bottom row.
\label{fig4}
}
\end{figure}
%%%%%%%%%%%%%%%%%%%%%%%%%%%%%%%%%%%%%%%%%%%%%%%%%%%%%%%%%%%%%%%%%%%%%%%%%%
\clearpage

\subsection{Methodology}

The classification scheme of 
the morphological types of galaxies is based on the RC3 classification
system \citep{deV91}, which is the de Vaucouleurs revised Hubble-Sandage 
system, or VRHS \citep{dev59}.
The description and sample color images given by \citet{but13} are also
consulted. The classification scheme is simplified by neglecting
stages of elliptical and lenticular galaxies. 
That is, the three stages of elliptical galaxies, compact ellipticals (cE), 
normal ellipticals (E), and late-type ellipticals (E$^{+}$) and 
the three stages of lenticular galaxies, S0$^{+}$, S0$^{0}$ and S0$^{-}$
are not distinguished. The reason for this is that
these E/S0 stages are based mostly on subtle aspects of structure
and ignoring them has little impact on our analysis. On the other hand, 
all of the stages of spiral galaxies
from 0/a to m are distinguished. dE/dSph galaxies
are separated from giant ellipticals because the surface brightness
distribution of dE galaxies is significantly 
different from that of the giant elliptical galaxies.

Elliptical galaxies are believed to have a simple structure. But, thanks to 
the  color information that is sensitive to the underlying 
stellar populations, it is possible to detect peculiar morphologies induced 
by recent star formation in elliptical galaxies. The discovery of blue
elliptical galaxies showing almost identical shapes to those of normal
elliptical galaxies but with different colors \citep{str01} is a good example
of the usefulness of color images in extragalactic astronomy.
Earlier studies of blue elliptical galaxies revealed anomalous 
blue cores \citep{mic99}.  For example. NGC 3156 
and NGC 4742 have blue cores which are caused by the presence of young
stellar populations in the nuclear regions \citep{mic99,suh10}. 

The distinction between elliptical galaxies and lenticular galaxies is made
mainly by their surface brightness distribution. Elliptical galaxies show 
steeper surface brightness gradients than
lenticular galaxies, particularly in the outer parts. This is why the
luminosity profiles of elliptical galaxies are best represented by the Sersic
function with $n\approx4$ whereas those of lenticular 
galaxies are best fit by the Sersic function with $n\approx1$. Galaxies with
characteristics intermediate between ellipticals and lenticulars are 
classified as E/S0. Lenticular galaxies are further divided into three bar
families, SA0, SAB0, and SB0, according to their bar types. The SA0 types
show no bar feature, whereas the SB0 types show a clear bar morphology crossing
the central bulge. Lenticular galaxies with intermediate bar types are 
assigned as SAB0. In the case of uncertain bar types, mostly due to high
inclinations, they are classified as S0 galaxies. 

For spiral galaxies, the spiral arm morphology is differentiated according to
the Hubble stages adopted in RC3 \citep{deV91}. Namely, 
the stages, 0/a, a, ab, b, bc, c, cd, d, dm, and m, of spiral galaxies
are determined according to the relative size of the bulge and the openness
of the spiral arms. On the other hand, the spiral varieties, 
(r) and (s), are not distinguished because they are less fundamental than 
stage and family, and, as for the E/S0 stages, ignoring them has little or
no impact on our analysis. We also exclude recognition of outer rings and
pseudorings for the same reason. Therefore, for example, a strongly barred
spiral galaxy with an intermediate Hubble stage is denoted as SBc, whereas
a non-barred spiral galaxy with an intermediate Hubble stage is denoted SAc.
For edge-on spirals, the bar families are not distinguished in this study
and the Hubble stage is denoted as Sa, Sab, Sb, etc.

Irregular galaxies are divided into three types: Im, dwarf irregular (dI), 
and BCD.
The first type, Im, is a natural extension of the late-type spirals and some
Im galaxies are believed to have a bar component. In that case, a morphological
type, IBm, is assigned. The second
type, dI, represents small dwarf galaxies with an amorphous shape.
Their physical properties, such as size, luminosity, and colors are similar to
those of dSph and dE$_{blue}$ galaxies, which are described below. 
The last type, 
BCD, stands for the blue compact dwarf and is characterized by star burst
regions. They have amorphous shapes with high central surface brightness, and
are likely to show an outer irregular envelope. However,
there is no unique definition of BCD in the literature \citep{kun00,gil03}
but the morphological properties of BCDs in the present sample are similar to
those of the BCDs analyzed by \citet{gil03}.
The shapes of BCD galaxies without an outer irregular envelope,
i.e., HII region-like BCDs, are similar to the small blue dwarf elliptical
galaxy (dE$_{blue}$) described below. BCDs show an extremely blue color due
to bursts of star formation. A significant fraction of irregular galaxies
have small star burst regions. These galaxies are distinguished from
Im and dI as well as BCD. We designate these galaxies as Ir/BCD. 

In addition to dI galaxies, there are five types of dwarf galaxies: dwarf 
lenticular galaxies (dS0), dE galaxies, blue-cored dwarf
elliptical galaxies (dE$_{bc}$), dSph galaxies, and dE$_{blue}$ galaxies.
We designate these five 
dwarf galaxies as dE-like galaxies. The dwarf lenticular
galaxies were introduced by \citet{san84} for the morphological
classification of dwarf galaxies in the Virgo cluster. They are
characterized by lens-like features that have a shallow brightness gradient 
interior to a sharp edge. (See Buta (2013) for further discussion of lenses in
disk galaxies.) In this respect, dS0 galaxies are different from their giant
cousins, S0 galaxies, which are characterized by a bulge and a disk. A majority
of dS0 galaxies show nucleation. The presence and absence of nucleation are
distinguished by the subscripts '$n$' and '$un$' as dS0$_{n}$ and dS0$_{un}$ 
for a nucleated one and an un-nucleated one, respectively.  Some dwarf galaxies
exhibit similar characteristics to dS0 galaxies but the outer parts of these
galaxies show disturbed disks, mostly resembling disrupted spiral arms. These 
galaxies are classified as dS0$_{p}$. We do not distinguish dS0$_{p}$ galaxies
from dS0 galaxies, except for the description of a detailed morphology.

dE galaxies can be distinguished from giant elliptical galaxies
by their surface brightness distribution, which is better represented by
the Sersic function with $n\approx2$.
Most previous studies do not distinguish between dE galaixes and
dSph galaxies (eg., Binggeli et al. 1985; Kormendy \& Bender 2012).
In particular, \citet{kor12} use 'Sph' to represent spheroidal galaxies that 
can be placed in a position parallel to the Im galaxies in their revised
parallel-sequence galaxy classification, which extends the van den Bergh
parallel-sequence galaxy classification \citep{ber76}. They assumed that the
Sph galaxies are stripped Scd-Im galaxies. However, our study considers 
dSph galaxies as a distinct population of dwarf galaxies because
photometric properties, such as surface brightness and colors of dSph
galaxies, seem to be different from those of dE
galaxies. Moreover, if we consider that dSph galaxies are mostly
dispersion-supported systems \citep{wal09,tol12} while a considerable fraction
of dE galaxies are supported by rotation \citep{geh10},
their origins might be different.

%%%%%%%%%%%%%%%%%%% Fig~5 (dwarf.eps (dEs.eps) ) %%%%%%%%%%%%%%%%%%%%
\begin{figure*}
%\tiny
\centering
\plotone{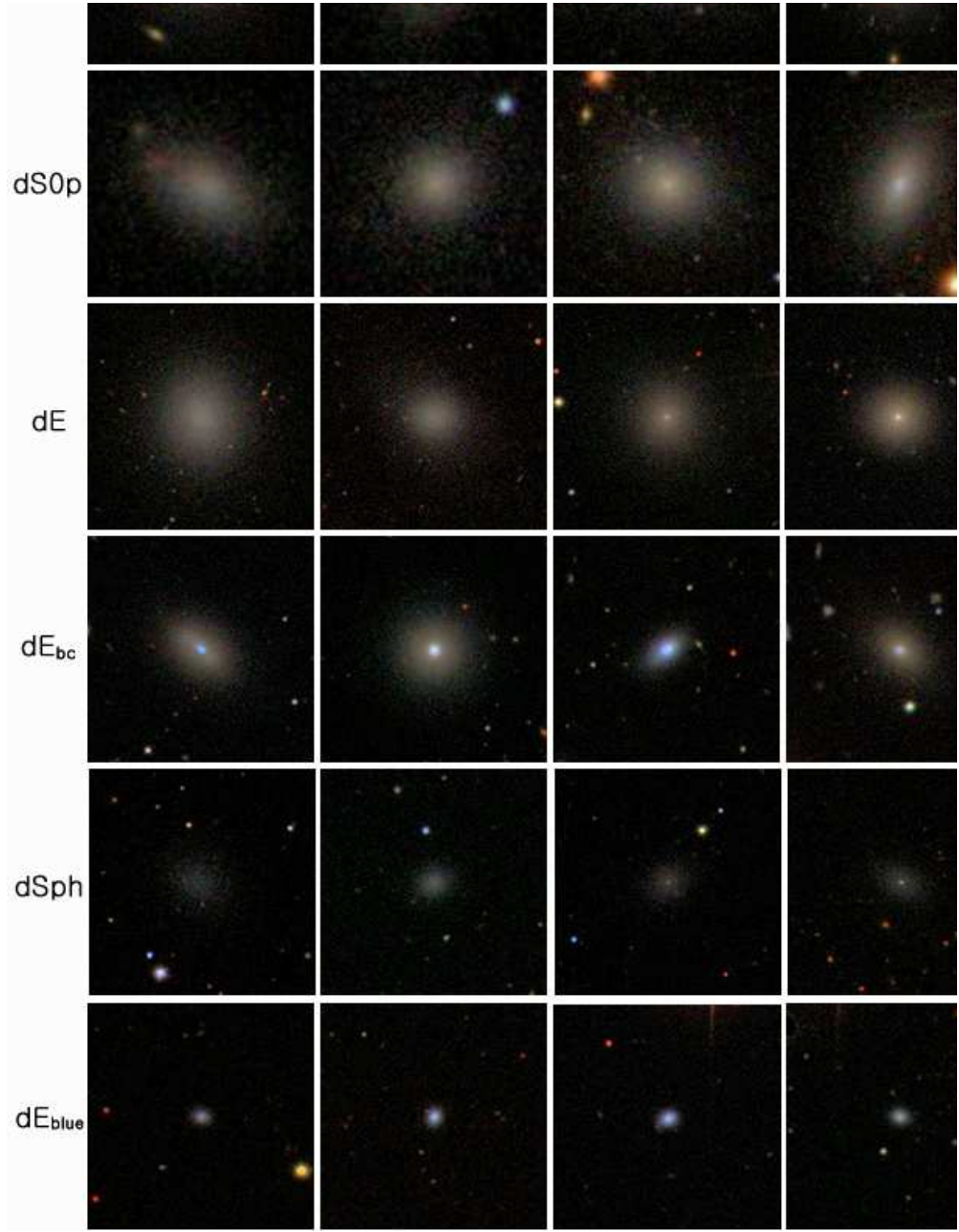}
\caption{Sample images for dwarf elliptical-like galaxies: dS0 (first row), 
dS0$_{p}$ (second row), dE (third row), dE$_{bc}$ (fourth row),  
dSph (fifth row), and dE$_{blue}$ (sixth row).
\label{fig5}
}
\end{figure*}
%%%%%%%%%%%%%%%%%%%%%%%%%%%%%%%%%%%%%%%%%%%%%%%%%%%%%%%%%%%%%%%%%%%%%%%%%%

The dE and dSph galaxies are distinguished by their surface brightness 
while dE$_{blue}$ galaxies are distinguished from others by
their blue color. The dE$_{bc}$ galaxies are characterized by the presence of 
blue cores which are supposed to be regions of active star formation. The
dE and dSph galaxies are further distinguished by their nuclear morphology:
dE$_{un}$, dE$_{n}$, dSph$_{un}$ and dSph$_{n}$. The distinction of nucleated
dwarfs from un-nulceated dwarfs is somewhat arbitrary because the degree of
nucleation varies continuously \citep{lis09}.
Blue-cored dwarfs, however, are easily distinguished
from nucleated dwarfs because the cores are bluer and larger than the
nuclear region of nucleated dwarfs. Blue dwarf ellipticals (dE$_{blue}$)
have similar color to irregular galaxies. Their size is similar to or smaller 
than HII region-like BCDs. Small dE$_{blue}$ and dI galaxies have similar 
features except for the axial symmetry.
On the other hand, there is no clear distinction
in the degree of axial symmetry for dE$_{blue}$. 
Therefore, it seems plausible that dE$_{blue}$ galaxies are 
dI galaxies with an extreme axial symmetry.

For practical purposes, the numerical type $T$ borrowed 
from RC3 is used with some simplification for the ellipticals ($T$=-5) and
lenticulars ($T$=-3). However, we keep all of the Hubble stages from T=0 to T=9 
for spirals (from $0/a$ to $m$) and $T$=10 (Im) and $T$=11 (dI) for irregulars. 
BCD is assigned as $T$=13, while irregulars containing small BCD-like components
as $T$=12. Also, we assign new codes from -6 to -11 for dwarf
elliptical-like galaxies: $T$=-6 (dE), $T$=-7 (dE$_{bc}$), 
$T$=-8 (dSph), $T$=-9 (dE$_{blue}$), 
$T$=-10 (dS0, dS0$_{p}$), and
$T$=-11 for transition type dwarf (dEs/dI dI/dEs), respectively.

In the present study, the color images of the sub-sample of galaxies were
examined using the image display tool provided by the SDSS for training
before the classification of the full sample of galaxies.
The morphological types given in the NED were consulted during the 
training period. Because the number of subtypes to be classified is more than 
20, morphological types were classified in two steps. Galaxies were divided
into four broad types (elliptical/lenticulars, spirals, irregulars, and
dE-like galaxies), and their fine subtypes were then classified after
sorting the galaxies according to their broad types. 
The entire sample of galaxies was classified by each of the three authors.
We followed two steps to finalize the morphological types of galaxies. In
the first step, one of the authors (Seo) determined the morphological types of
the entire sample by consulting the results of other authors. The final morphological
type of each galaxy was determined by the senior author after reviewing 
Seo's classification. There is not much difference between the final
morphological types and those of Seo for giant galaxies (E, S0, and Sp) and 
irregular galaxies. However, about $10\%$ of Seo's classification of dE-like
galaxies are revised, mostly within the category of dE-like galaxies. \\

\subsection{Type Examples}

Figure~4 shows representative images of elliptical, lenticular, spiral,
and irregular galaxies classified in the present study. The elliptical and
lenticular galaxies are displayed in the first and second rows, respectively. 
Spiral galaxies are presented in the third, fourth, and fifth rows, and are
grouped according to the bar types (SA, SAB, and SB). The spiral galaxies
are arranged according to the Hubble stage (a, b, c, d, and m)
with the earliest types (SAa, SABa, and SBa) in the leftmost column and the
latest types (SAm, SABm, and SBm) in the rightmost column. The irregular
galaxies are presented in the sixth row. The first three columns present
Im, dI, and Ir/BCD galaxies, respectively, whereas the fourth column and 
the fifth column display a BCD embedded in an irregular envelope and
a BCD resembling HII regions, respectively.
Different scales are used for each stamp image in Figure 4 for a better 
demonstration of their morphology.

%%%%%%%%%%%%%%%%%%%%%%%%%%%%%%%%%%%%%%%%%%%%%%%%%%%%%%%%%%%%%%%%%%%%%
%%%%%%%%%%%%%% Fig~6 (HTbroad.ps) %%%%%%%%%%%%%%%%%%%%
\begin{figure}
\epsscale{.80}
\plotone{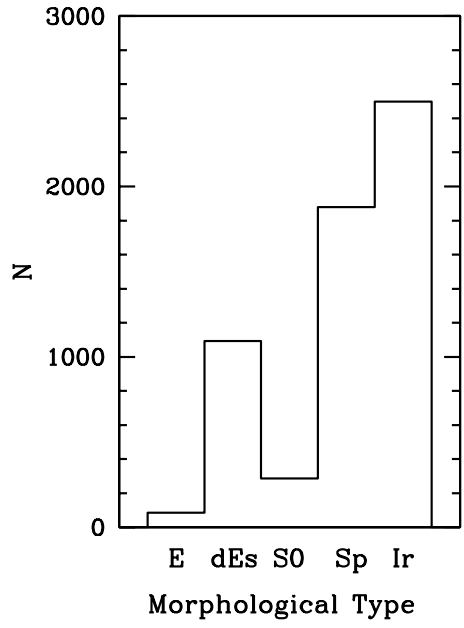}
\caption{Morphology frequency of the 5836 galaxies within $z = 0.01$.
\label{fig6}
}
\end{figure}
%%%%%%%%%%%%%%%%%%%%%%%%%%%%%%%%%%%%%%%%%%%%%%%%%%%%%%%%%%%%%%%%%%%%%

\clearpage

Figure~5 shows images of dE-like galaxies. The first and second rows
display images of dwarf lenticular galaxies, with normal
dwarf lenticulars (dS0) in the first row and dwarf lenticular galaxies with
peculiar disk morphology (dS0$_{p}$) in the second row. The morphology of
dS0 galaxies is characterized by the presence of a lens-like
feature in addition to the disk component. Lens-like features are normally
redder than the disk but some dS0 galaxies have blue lens-like features 
due to young stellar populations near the nuclei. The
galaxies in the third rows are dE galaxies. We arranged dE$_{un}$ galaxies
in the first and second columns and dE$_{n}$ galaxies from third to fifth
columns. In the fourth row, we display dE$_{bc}$ galaxies that show  
intense star formation in the nuclei of the galaxies. The nuclear star
formation is thought to be caused by recent gas accretion \citep{san08,kre11}.
The fifth and sixth rows show dSph and dE$_{blue}$, respectively
The dSph galaxies are generally
fainter and bluer than the dE galaxies.
Some dE$_{blue}$ galaxies have large blue cores, the radius of which is
approximately half the galaxy size.

In addition to these types of galaxies, there are compact elliptical galaxies
that are occasionally observed near the centers of clusters.
They are smaller than the giant elliptical galaxies but their central
surface brightness is similar to that of giant elliptical galaxies.
Due to their high surface brightness, they are classified as compact
elliptical galaxies rather than dE galaxies.
A prototype of compact elliptical galaxies is M32 in the LG.

\section{Morphology Catalog of Local Galaxies}
This section presents a catalog of the visually classified morphological 
types of 5836 galaxies whose redshifts are less than $z=0.01$.
The basic statistics of the morphological types of the local galaxies are
reported along with a brief comparison with other studies. In addition,
the new morphological types revealed from color images are briefly described.

\subsection{Catalog}

The morphological types of 5836 galaxies are presented along with some basic 
data obtained from the SDSS DR7. Table 1 lists a part of the Morphological 
Catalog of Local Galaxies whose full content is available in the electronic
version of this paper. The SDSS object identification number is used as the 
name of a galaxy but the primary NED name is presented in the last column 
for reference. This paper uses the distance of a galaxy derived from the
heliocentric redshift corrected for the motion of the LG. We used the
prescription of \citet{mou00} to derive the velocity relative to the centroid
of the LG ($V_{\rm{LG}}$). On the other hand,
for those galaxies lying inside a $10^{\circ}$-cone around M87
with a heliocentric redshift less than z=0.007 \citep{kra86}, the 
distance of the Virgo cluster is used. 
A Hubble constant of H=75kms$^{-1}$ was used.
The catalog contains 12 columns with the following information.

\indent{Column 1: SDSS Object ID. } \\
\indent{Column 2: Right ascension (J2000.0) in degrees.} \\
\indent{Column 3: Declination (J2000.0) in degrees.} \\
\indent{Column 4: Heliocentric redshift determined by spectroscopic 
observations} \\
\indent{Column 5: Morphological types determined in the present study.}\\
\indent{Column 6: Morphological index $T$. The $T$ code is adopted from RC3 but
somewhat simplified except for spirals. We add new codes to distinguish
subtypes of dwarf galaxies (see section 3.1 for the decription of $T$ code).}\\
\indent{Column 7: Distance in Mpc derived from the radial velocity relative to
the LG.} \\
\indent{Column 8: Absolute magnitude in the $r$-band ($M_{r}$). We use the
model magnitude from SDSS DR7} corrected for the galactic extinction \citep{sch98}. The SDSS model magnitude is a magnitude derived by
best-fitting model (exponential profile or de Vaucouleurs profile) to the 
observed images. \\
\indent{Column 9: Model $u-r$ color (extinction corrected)} \\
\indent{Column 10: Isophotal semi-major axis in pixels measured at 
$\mu_{r}=25$ mag arcsec$^{-2}$ } \\
\indent{Column 11: Isophotal axis-ratio} \\
\indent{Column 12: Galaxy names in the literature. The first entry
in NED is taken. In cases of no NED name, we make a name using the 
coordinates of the galaxy similar to the SDSS object name.} \\

\clearpage

\begin{deluxetable}{ccrrrrrrrrcr}
\tabletypesize{\scriptsize}
\rotate
\tablecaption{Catalog of the morphological types of local galaxies.\label{tbl-1}}
\tablewidth{0pt}
\tablehead{
\colhead{ID} & \colhead{RA} & \colhead{DEC} & \colhead{z} & \colhead{Morphology} &
\colhead{T} & \colhead{D (Mpc)} & \colhead{M$_{r}$} &
\colhead{$u-r$} & \colhead{$a_{iso}$} &
\colhead{$(b/a)_{iso}$} & \colhead{name}
}
\startdata
587731514215760014  &    17.283075   &   1.121029   &    0.003855   &    dI  &  10  &   17.5 & -14.79 & 0.73 &  44.7 &  0.5  &  SHOC053  \\
588015510351118339  &   18.232364    &  0.981506    &   0.003843    &   SABdm & 8    &  17.43 & -18.79 & 1.86 &  -9   &  -9   &  NGC0428  \\
588015510351184046  &   18.413968    &  0.875452    &   0.003873    &   Sm/Im & 10   &  17.55 & -13.73 & 0.69 &  26.2 &  0.71 &  UGC00772 \\
588015510351249607  &   18.584742    &  0.916687    &   0.003717    &   dI   & 10   &  16.92 & -14.77 & 1.05 &  41.4 &  0.5  &  SDSSJ011420.41+005503.2 \\
588015507666960497  &   18.719523    &  -1.095353   &   0.006761    &   dS0n  & -10  &  29.05 & -16.02 & 1.45 &  40.4 &  0.62 &  SDSSJ011452.68-010543.2 \\
587743981974650974  &   18.84512     &  8.099593    &   0.007812    &   SAcd  & 6    &  33.24 & -17.09 & 2.39 &  61.4 &  0.74 &  UGC00803  \\
587731511532060697  &   18.876839    &  -0.86098    &   0.005874    &   SABcd & 6    &  25.51 & -18.95 & 1.54 &  149.1&  0.77 &  NGC0450  \\
587731511532191838  &   19.087214    &  -0.911365   &   0.005576    &   dI   & 10   &  24.32 & -15.12 & 1.07 &  26.9 &  0.85 &  SDSSJ011620.92-005440.8 \\
758877280840910682  &   19.106185    &  33.431996   &   -0.000627   &   dSphun& -8   &   0.64 & -10.52 & 2.62 &  4.7  &  0.74 &  AndII  \\
758877278149148728  &   19.18239     &  16.400064   &   0.006615    &   SAa   & 1    &  28.46 & -18.74 & 1.74 &  88.2 &  0.76 &  CGCG459-021 \\
587727226772258858  &   19.377137    &  -9.296768   &   0.006479    &   SAm   & 9    &  27.91 & -16.47 & 1.11 &  73.9 &  0.37 &  SDSSJ011730.51-091748.3 \\
587727179531747451  &   19.809026    &  -9.597428   &   0.006379    &   Im/BCD& 12   &  27.5  & -14.81 & 1.16 &  29.5 &  0.43 &  SHOC061 \\
587731512606326893  &   20.028347    &  -0.205448   &   0.00581     &   Sdm   &8     & 25.23  & -17.18 & 1.52 &  106.6&  0.29 &  UGC00866  \\
588015510351970406  &   20.26232     &  0.851495    &   0.008134    &   dI   & 10   &  34.48 & -15.83 & 1.02 &  24.4 &  0.69 &  SDSSJ012102.95+005105.3 \\
587744045327319127  &   20.364876    &  7.018498    &   0.007495    &   Sa    & 1    &  31.94 & -19.3  & 1.65 &  111  &  0.39 &  NGC0485  \\
588015509278359598  &   20.522909    &  0.08518     &   0.007569    &   SAdm  & 8    &  32.23 & -15.84 & 1.58 &  77.8 &  0.14 &  FGC0156  \\
588015510352101409  &   20.537435    &  0.945465    &   0.007799    &   Scd   & 6    &  33.14 & -19.91 & 1.92 &  310.6&  0.19 &  NGC0493  \\
588015508204748937  &   20.810885    &  -0.700637   &   0.006625    &   Sd    & 7    &  28.46 & -16.83 & 1.38 &  88.3 &  0.23 &  UGC00931  \\
587743980364955756  &   20.93191     &  6.691629    &   0.007789    &   SABcd & 6    &  33.09 & -17.14 & 1.74 &  50.7 &  0.96 &  UGC00942  \\
\enddata
%% Text for table notes should follow after the \enddata but before
%% the \end{deluxetable}. Make sure there is at least one \tablenotemark
%% in the table for each \tablenotetext.
\tablecomments{Table \ref{tbl-1} is published in its entirety in the
electronic edition of the {\it Astrophysical Journal Supplement}. A randomly
selected portion is shown here. Color images of the galaxies listed in Table 1
are provided in the web (http://earth.es.pusan.ac.kr/hbann/apjs/z01image).}
\end{deluxetable}

%%%%%%%%%%%%%% Fig~7 (Hsp.ps) %%%%%%%%%%%%%%%%%%%%
\begin{figure}
\epsscale{.80}
\plotone{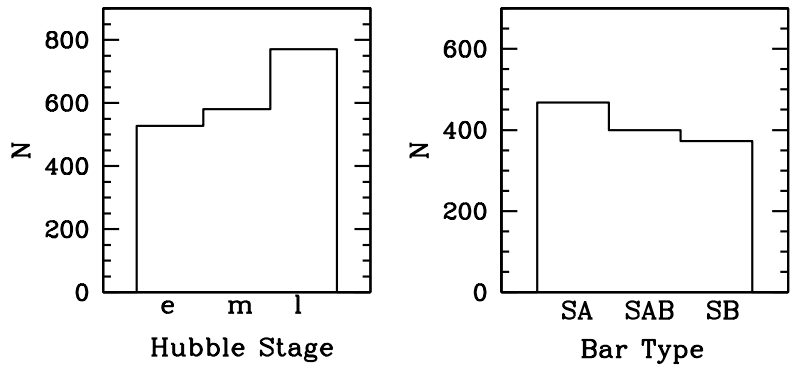}
\caption{Frequency of spiral galaxies as a function of the Hubble
stage (early (e), intermediate (m), and late (l)) in the left and 
frequency of spiral galaxies as a function
of the bar type (SA, SAB, and SB) in the right. \label{fig7}
}
\end{figure}
%%%%%%%%%%%%%%%%%%%%%%%%%%%%%%%%%%%%%%%%%%%%%%%%%%%%%%%%%%%%%%%%%%%%%

\clearpage

%%%%%%%%%%%%%% Fig~8 (HdESph.ps) %%%%%%%%%%%%%%%%%%%%
\begin{figure}
\plotone{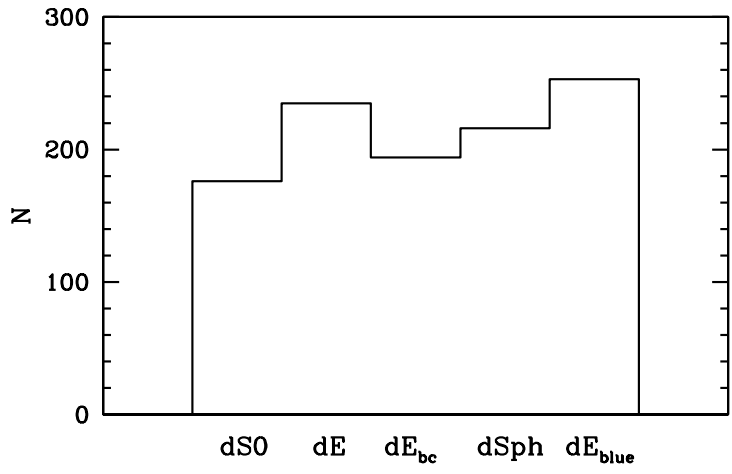}
\caption{Frequency of subtypes of dwarf elliptical-like 
galaxies (dEs). 
\label{fig8}
}
\end{figure}
%%%%%%%%%%%%%%%%%%%%%%%%%%%%%%%%%%%%%%%%%%%%%%%%%%%%%%%%%%%%%%%%%%%%%
\clearpage

\begin{deluxetable}{rrrrrr}
\tabletypesize{\scriptsize}
\tablecaption{Frequency of the morphological types of local galaxie \label{tbl-2}}
\tablewidth{0pt}
\tablehead{
\colhead{ } & \colhead{E} & \colhead{dEs} & \colhead{S0} & \colhead{Sp} &
\colhead{Ir}
}
\startdata
number &   86 &   1092 &    286 &   1878 &   2498 \\
fraction & 0.015 &   0.187 &   0.049 &   0.321 &   0.428 \\
\enddata
\end{deluxetable}

%%%%%%%%%%%%%% Fig~9 (HFdEur2.ps  Hurdwarf.dat) %%%%%%%%%%%%%%%%
\begin{figure}[!!t]
\epsscale{.80}
\plotone{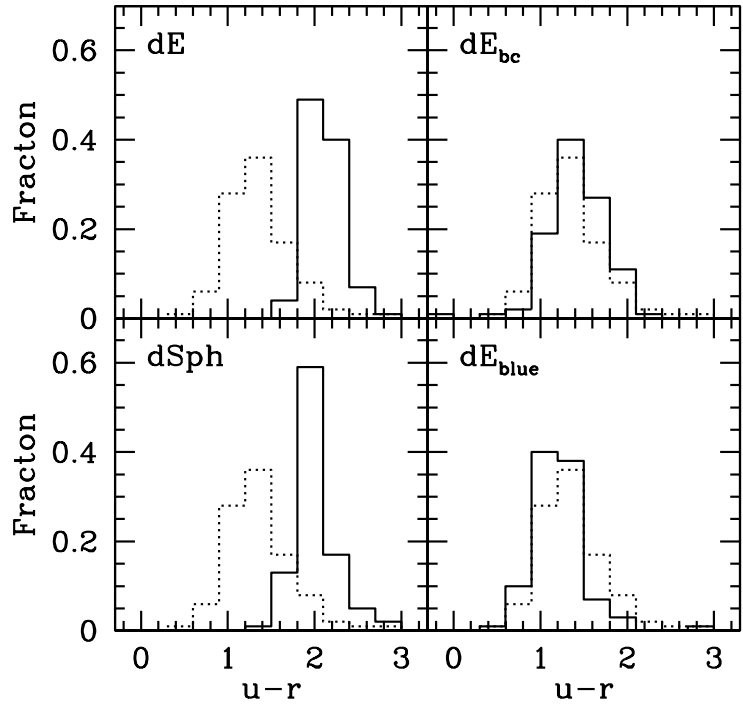}
\caption{Color distributions of dwarf galaxies: dE, dE$_{bc}$, dSph and
dE$_{blue}$. The histograms plotted by dotted lines represent dI 
galaxies. \label{fig9}
}
\end{figure}
%%%%%%%%%%%%%%%%%%%%%%%%%%%%%%%%%%%%%%%%%%%%%%%%%%%%%%%%%%%%%%%%%%%%%

\clearpage

\subsection{Frequency Distributions of Morphological Types}

Figure 6 presents the frequency distribution of the broad morphological 
types of 5836 galaxies. Irregular galaxies are most abundant ($42.8\%$),
and spiral galaxies are second-most abundant ($32.1\%$).
These two types comprise $\sim75\%$ of the local galaxies.
The fractions of elliptical galaxies and lenticular galaxies are $0.015$
and $0.049$, respectively. The fraction of dE-like galaxies
is $0.187$. If we exclude the dS0 galaxies, it becomes $\sim0.15$. 
Therefore, they are 10 times more frequent than their giant
cousins, elliptical galaxies. The local universe is dominated by dwarf galaxies
because the majority of irregular galaxies are dwarf galaxies.
Table 2 lists the basic statistics of the morphological types of galaxies
presented in this catalog (Table 1).

Figure 7 shows the frequency of the Hubble stages (top panel) and
that of the bar family (bottom panel). The Hubble stages are grouped
into three stages of early (e), intermediate (m), and late (l).
The edge-on galaxies are included in the frequency distribution
of the Hubble stages but they are not included in the frequency 
distributions of the bar family. The number of spirals increases from the
early types to the late types.
The fraction of barred galaxies (SAB and SB) is $\sim0.62$ which is similar 
to the bar fraction in RC3 \citep{deV91} and that derived from the
mid-IR images of nearby galaxies \citep{but15} 
but somewhat higher than those derived from the SDSS galaxies at $z > 0.01$
\citep{lee12,ohs12,ski12}. 
The bar fraction from the ellipse fitting technique (e.g., Barazza et al. 2009)
is somewhat lower than the visual classifications but that from
Fourier analysis \citep{kra12} is somewhat higher
than that in the present study. The difference in the bar fractions is due
mostly to the classification method and the criteria for bar selection. 
The lowest bar fraction from the SDSS data is that derived by
\citet{ohs12} who reported a bar fraction of $0.36$.

Figure 8 shows the frequency distribution of the subtypes of dE-like 
galaxies; dE, dE$_{bc}$, dSph, dE$_{blue}$, and
dS0. As shown in Figure 8, the dominant population of the dE-like 
galaxies is dE$_{blue}$ galaxies, which is distinguished
from the others by its blue color. The second largest population is dSph 
galaxies, but if dE$_{bc}$ galaxies are considered to be a special type of 
dE, then the number
of dE is similar to that for dSph. The dS0 galaxies
include dS0$_{p}$ galaxies which show signatures of spiral arm
structures in the outer parts.

Figure 9 shows the frequency distributions of $u-r$ colors for the subtypes 
of dE-like galaxies with those of dI 
for comparisons. The dE galaxies show the reddest color with 
narrow color ranges, whereas the 
dE$_{blue}$ galaxies show the bluest color similar to that of dI galaxies.
The dSph galaxies show redder colors than dE$_{bc}$ and dE$_{blue}$. Their
average color is similar to that of dE galaxies but they have a u-r 
color distribution extended to blue colors up to $u-r\approx 1$.

\subsection{Comparison with Others}

This classification is compared with the automated classification
\citep{par05} given in KIAS-VAGC. Because they presented broad morphological
types (early and late types) only,  the types in this study were also 
grouped into early and late types.
The early type includes ellipticals (E) and lenticulars (S0) with their dwarf
cousins (dE, dE$_{bc}$, dSph, dE$_{blue}$, and dS0) and the late type includes 
spirals (Sp) and irregulars (Ir). Spiral galaxies show the best
match ($94\%$) and dE-like galaxies show the worst match ($44\%$)
between the present classification and that reported by \citet{par05}. 
Ellipticals (E) and irregulars (Ir) also show good matches of 
$\sim90\%$, whereas lenticulars show fair matches ($73\%$).
The low match between the present classification and \citet{par05} for dE-like 
galaxies is due mostly to the extremely high mismatch of 
dE (dE$_{un}$ and dE$_{n}$) and dSph (dSph$_{un}$ and dSph$_{n}$), which show
only $30\%$ matches. For both dE and dSph, the un-nucleated ones show better
matches. The worst match was found for nucleated dwarf spheroidals (dSph$_{n}$)
which showed only a $24\%$ match. 
On the other hand, dE$_{bc}$ and dE$_{blue}$ show the $\sim60\%$ matches.
The reason for the worse classification of dE and dSph galaxies appears to be
the small number of dwarf ellipticals in the training sample of \citet{par05}.
In the case of dE$_{bc}$ and dE$_{blue}$ galaxies, they included a sufficient 
number in their training sample. 

%%%%%%%%%%%%%% Fig~10 (nedcomp3.ps) %%%%%%%%%%%%%%%%%%%%
\begin{figure}
\epsscale{.80}
\plotone{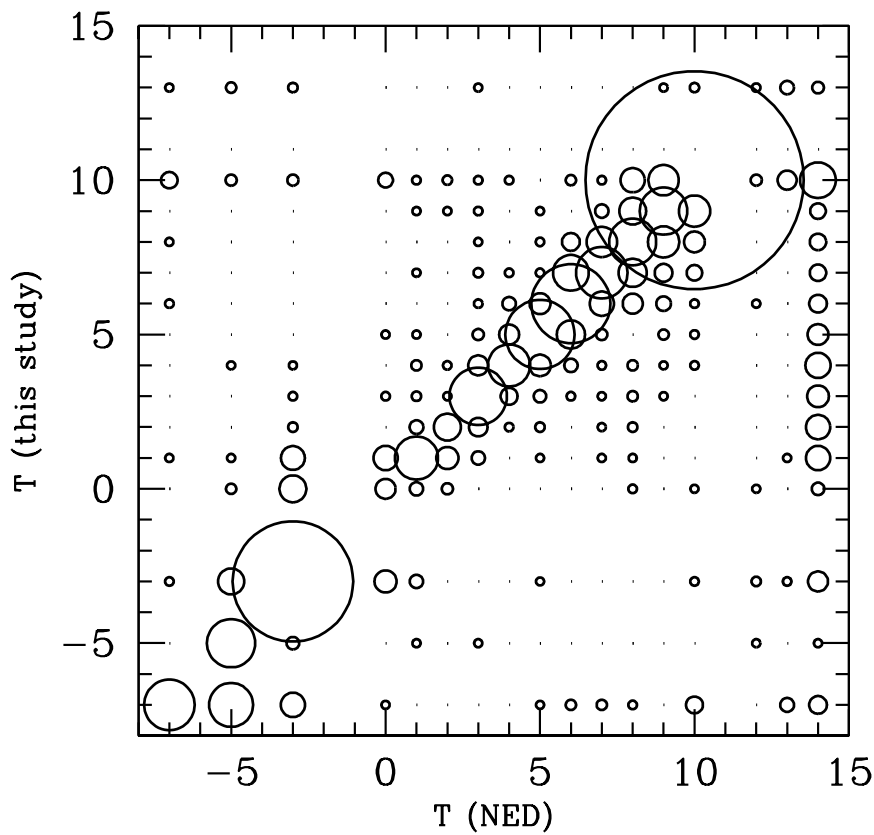}
\caption{Comparisons of morphological types determined by the present study 
and those in the NED. We use the same numerical code T as those in RC3 for
spiral galaxies ($T$=0 to $T$=9). Ellipticals and lenitculars are coded 
as $T$=-5 and $T$=-3, respectively.
The dwarf elliptical-like galaxies are coded as $T$=-7
and irregulars (Im, dI, and Ir/BCD) are coded as $T$=10 and BCDs as $T$=13. 
In addition,
$T$=12 is assigned to I0 galaxies and $T$=14 for spirals with unknown stages.
\label{fig10}
}
\end{figure}
%%%%%%%%%%%%%%%%%%%%%%%%%%%%%%%%%%%%%%%%%%%%%%%%%%%%%%%%%%%%%%%%%%%%%

\clearpage

Figure 10 compares the morphological types determined in this study with
those from NED. We use the numerical type $T$ adopted in the present catalog 
with three additional codes: $T$=12, $T$=13, and $T$=14. The first and second ones
are used for the NED I0 galaxies and BCD galaxies, respectively. The last one
is assigned to the spiral galaxies with no Hubble stage. They
are usually given as S? or SB?, which suggests an uncertain morphology. In 
addition to these types, $T$=-7 is assigned to dE-like galaxies.
The circle sizes in Figure 10 are
proportional to the fractions of the types classified in the present study. As 
shown in Figure 10, there is good agreement between the present types and those
from the NED. On the other hand, there are two types that show a considerable 
difference. One type is elliptical galaxies ($T$=-5) and 
the other is I0 galaxies ($T$=12).
Some bias in the classification of S0/a galaxies can be seen.
The NED S0/a galaxies ($T$=0) are classified mainly as one of the
three types, S0, S0/a, and Sa. If the S0/a is highest, as in Figure~10, 
then the Sa fraction would be slightly higher than the others.

For elliptical galaxies, approximately half of the NED ellipticals are 
classified as lenticulars or dE-like galaxies in the present
study. This is expected because there is frequent
confusion between elliptical and lenticular galaxies 
in the literature. Note that most NED lenticular galaxies are classified as 
lenticular galaxies, whereas NED elliptical galaxies are classified
frequently as lenticular galaxies. The confusion between ellipticals
and dE-like galaxies is also conceivable
because they differ only in the surface brightness and its gradient. 
As shown in Figure 10, ellipticals are more likely to be misclassified as dwarf
ellipticals/spheroidals than lenticulars. 

I0 is a special type of irregular galaxy.  It is not so
irregular because it is not as disorganized, which is the
main characteristics of irregular galaxies. I0 galaxies show a smooth structure 
similar to lenticular galaxies but have central irregular zones of active
star formation \citep{but13}. This type is not included as a separate class
because they do not have properties that can be easily distinguishable from
other types of galaxies. Recent observations show that many 
elliptical/lenticular galaxies show similar features in their central regions.
As expected, most I0 galaxies are classified in the present study as either
lenticular galaxies ($T$=-3) or irregular galaxies ($T$=10).
\textbf{In the mid-IR, I0s like NGC 2968, 3077, and 5195 are much more regular
and the class is no longer applicable \citep{but15}.}

%%%%%%%%%%%%%% Fig~11 (RC3.ps) %%%%%%%%%%%%%%%%%%%%
\begin{figure}
\epsscale{.80}
\plotone{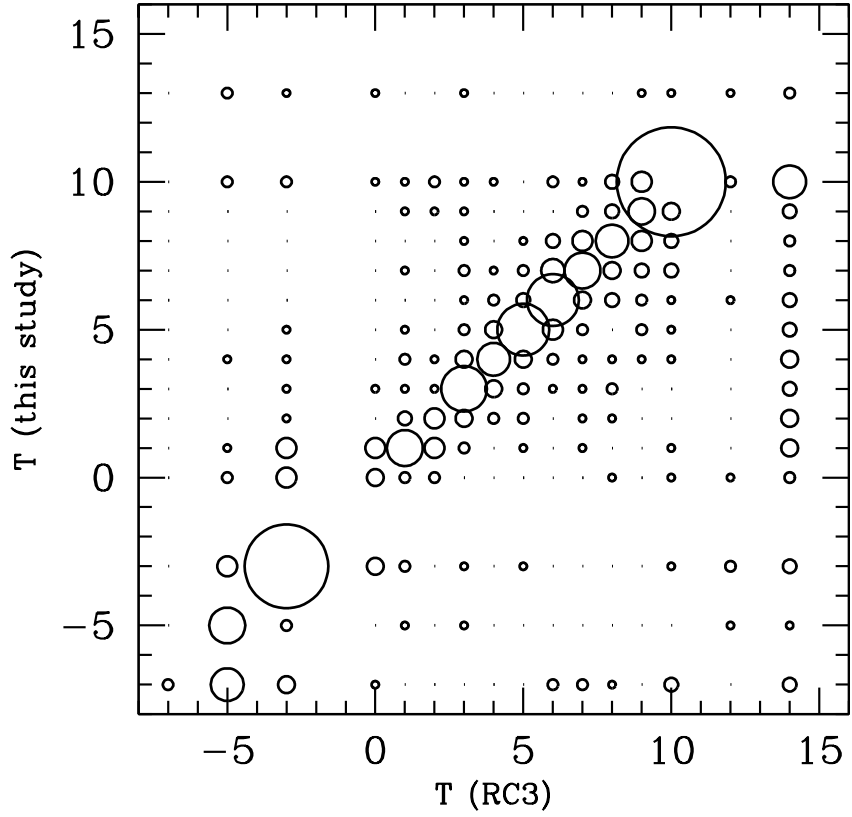}
\caption{Comparisons of morphological types determined by the present study
and those in the RC3. The $T$ codes are the same as those in Figure 10. 
\label{fig11}
}
\end{figure}
%%%%%%%%%%%%%%%%%%%%%%%%%%%%%%%%%%%%%%%%%%%%%%%%%%%%%%%%%%%%%%%%%%%%%

A considerable difference between the present types and NED types can also be
observed in BCD galaxies ($T$=13). Most of our BCDs are also classified as 
BCDs in NED.
However, the majority of NED BCDs are classfied as irregulars ($T$=10). This 
is due to those Ir/BCD galaxies which are considered as irregular galaxies in
our classification but most of which are classified as BCDs in NED. 
Confusion with irregular galaxies is natural because many irregular galaxies
have star burst regions, which is a basic characteristic of BCDs.
Such irregular galaxies are considered to be BCDs in the present study only
if the star burst regions are large enough to dominate the luminosities of 
galaxies. The
introduction of dE$_{blue}$ galaxies is another cause of the large mismatch
between the NED BCDs and our types for these galaxies. Similar fractions of
NED BCDs are classified as BCDs and dE$_{blue}$ galaxies in our classification.
Some dE$_{blue}$ galaxies are small enough, i.e., comparable to 
HII region-like BCDs with similarly blue color with the BCD color.

The circles at $T$=-7 represent early-type dwarf galaxies. They are not divided 
into subtypes of dE, dE$_{bc}$, dE$_{blue}$, dSph, and dS0 because the
NED types were not finely defined. Most NED early-type
dwarfs are classified by us as early-type dwarfs but a small fraction of 
NED early type dwarfs are classified as irregular galaxies. A better
comparison of early-type dwarfs with others is given below (Figure 13). 
Finally, the small circles at $T$=14 represent spiral galaxies with no Hubble
stages in NED. Most spiral galaxies we classify as spirals
but some are classified as lenticulars, early-type dwarfs, irregulars,
and BCDs.

%%%%%%%%%%%%%% Fig~12 (VCC.ps) %%%%%%%%%%%%%%%%%%%%
\begin{figure}
\epsscale{.80}
\plotone{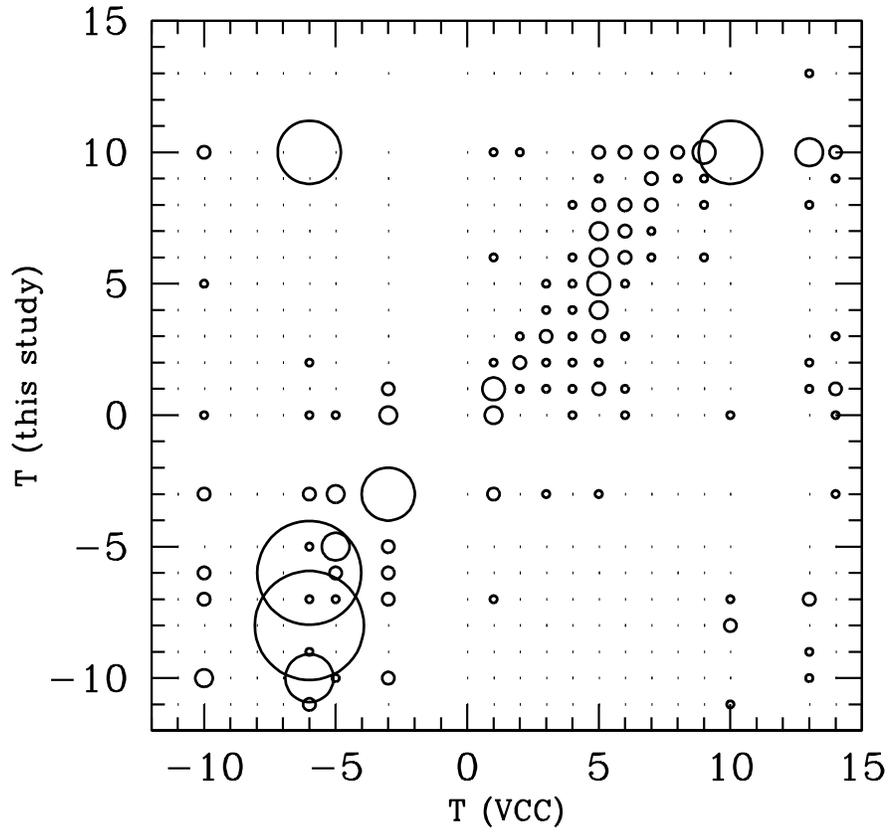}
\caption{Comparisons of morphological types determined by the present study and
those in the VCC. The $T$ (VCC) codes are the same as $T$ (NED) in Figure 10  
except for $T$ (VCC)=-6 and $T$ (VCC)=-10 which are assigned to dEs and dS0,
respectively.
\label{fig12}
}
\end{figure}
%%%%%%%%%%%%%%%%%%%%%%%%%%%%%%%%%%%%%%%%%%%%%%%%%%%%%%%%%%%%%%%%%%%%%

Figure 11 compares  the present types with those of RC3 \citep{deV91}.
We used the same convention of $T$ type. 
As shown in Figure 10 and Figure 11, the two comparisons
are very similar. This similarity is due to the fact that most of the NED types
come from RC3. For the galaxies in the present catalog, $95\%$ of NED types
come from RC3. The absence of galaxies at $T$ (RC3)=13 is due to the lack of 
BCD types in RC3. There are 18 galaxies whose morphological types are 
classified as peculiar galaxies (P) in RC3. The majority of these galaxies 
are classified as irregulars (7 galaxies) and BCDs (5 galaxies). The remaining
6 galaxies consist of four spirals, one lenticular galaxy, and one dE
galaxy.

Figure 12 shows the comparison of the morphological types with those in 
the Virgo Cluster Catalog (Binggeli et al. 1985, hereafter VCC).
Since in VCC, dS0 galaxies are distiguished from other dE-like
galaxies, we assign $T$ (VCC)=-6 to dEs and $T$ (VCC)=-10 to dS0 in Figure 12.
However, we keep our subtypes of dE (-6), dE$_{bc}$ (-7), dSph (-8), 
dE$_{blue}$ (-9), dS0 (-10), and dEs/dI (-11) in Figure 12 to see how VCC dEs
distribute in our dE-like galaxies. For the types
from $T$=-5 to $T$=14, we used the same conventions as those in Figure 11. As
shown in Figure 12, there is a fairly good agreement between the morphological 
types determined in this study and those in VCC.  However, the degree of 
agreement is worse than RC3, especially for the Hubble stages of spiral 
galaxies later than $T$ (VCC)=3. The best agreement between this study and VCC
is observed for lenticular galaxies. Elliptical galaxies also give quite
good agreement. On the other hand, the greatest discrepancy between this study
and VCC is observed for VCC BCDs. The majority of VCC BCDs are classified as
irregulars in this study. There are two dwarf spiral galaxies in VCC. One is
IC 3094, which is given a type of $\lq$dS?$\rq$. The NED type of this galaxy 
is $\lq$S; BCD$\rq$ and we classified it as $\lq$SAa$\rq$. 
The other is NGC 4192 which is  calssified as $\lq$d:Sc$\rq$
in VCC. We classified it as $\lq$SAcd$\rq$ while NED gives 
$\lq$SAB(s)ab$\rq$.

%%%%%%%%%%%%%% Fig~13 kmk14.ps %%%%%%%%%%%%%%%%%%%%
\begin{figure}
\epsscale{.80}
\plotone{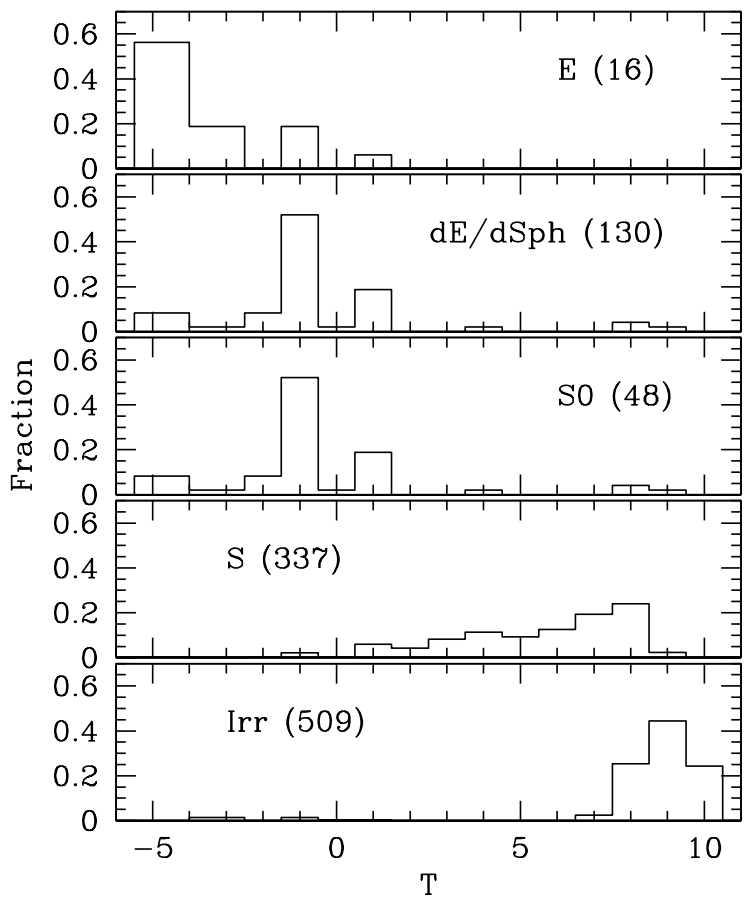}
\caption{Comparison of present morphological types and those in the KN.
We use the numerical code $T$ of \citet{kar13a}: $T$=-3 (dSph), $T$=-2 (dE),
$T$=-1 (S0), and $T$=0 (dS0). From $T$=1 to $T$=8 are assigned to spirals 
as $T$=1 for Sa and $T$=8 for Sdm and Sm.
Irregulars including BCDs are assigned as $T$=9 
and transition types as $T$=10.
\label{fig13}
}
\end{figure}
%%%%%%%%%%%%%%%%%%%%%%%%%%%%%%%%%%%%%%%%%%%%%%%%%%%%%%%%%%%%%%%%%%%%%
\clearpage

Figure~13 compares the types in this study with those reported by
\citet{kar11,kar13a,kar13b,kar13c,kar14}. These studies (KN hereafter) 
include the Updated Nearby Galaxy Catalog.
We plot the fractional distributions of the morphological types of
galaxies in the KN that are listed in our catalog. Our morphological
types are grouped into five broad types E, dEs, S0, Sp, and Ir, 
and we display the fractional frequency distributions of $T$ in KN at
each panel, which are labeled according to morphological type with 
the number of galaxies in parentheses. The $T$ codes adopted by \citet{kar13a},
which are slightly different from the $T$ codes used in the present catalog, 
are used for convenience.

The elliptical galaxies are presented in the top panel. There are only 16 
shared galaxies, 9 of which are classified as ellipticals in KN while all
of the other galaxies except one (MCG-01-33-007) are classified
as dEs or lenticulars. The type $T$=-2 and $T$=-3 were assigned for dE 
and dSph, respectively, in KN.
The second panel shows the fractional distribution of the morphological types
of galaxies classified as dE-like galaxies in our
catalog. Our dwarf galaxies designated as dEs include dE,
dE$_{bc}$, dSph, dE$_{blue}$, and dS0s. Here, we present
here 131 dEs that are shared with KN. As shown in the second panel
of Figure 13, there is a significant disagreement in the morphological types
between ours and theose of KN. Among 130 dEs, 18 galaxies are
classified as irregular galaxies and 42 galaxies as spirals, which are mostly
late-types of stages dm and m ($T$=8). 
The third panel from the top in Figure 13 shows 48 lenticulars listed in the
present catalog, which show a wide range of $T$ in \citet{kar13a}. About $53\%$
of our lenticular galaxies are classified as $T$=-1, which is assumed 
to designate lenticular galaxies in KN.
Among the 22 galaxies that were not classified
as lenticulars in KN, 4 galaxies are ellipticals and 5 are dEs.
In particular, 10 galaxies that correspond to $20\%$ of the S0 galaxies are 
classified as early-type spirals of stage $0/a$ and $a$.
The fourth panel from the top shows the fractional distribution of the
KN morphological types for our spiral galaxies. There is a good agreement
between the current classification and those in KN. The fraction of
spirals with $T > 4$ is $65\%$, which is in good agreement with that of the
present catalog ($64\%$). As shown in the bottom panel of Figure 13, most of
irregular galaxies ($\sim95\%$) are classified as $T$=8, $T$=9, 
and $T$=10 in the KN where late-type spiral galaxies (dm and m),
irregulars including BCDs, and transition types are classified as $T$=8,
$T$=9, and $T$=10, respectively.

%%%%%%%%%%%%%% Fig~14 Mrur.ps %%%%%%%%%%%%%%%%%%%
\begin{figure}
\epsscale{.80}
\plotone{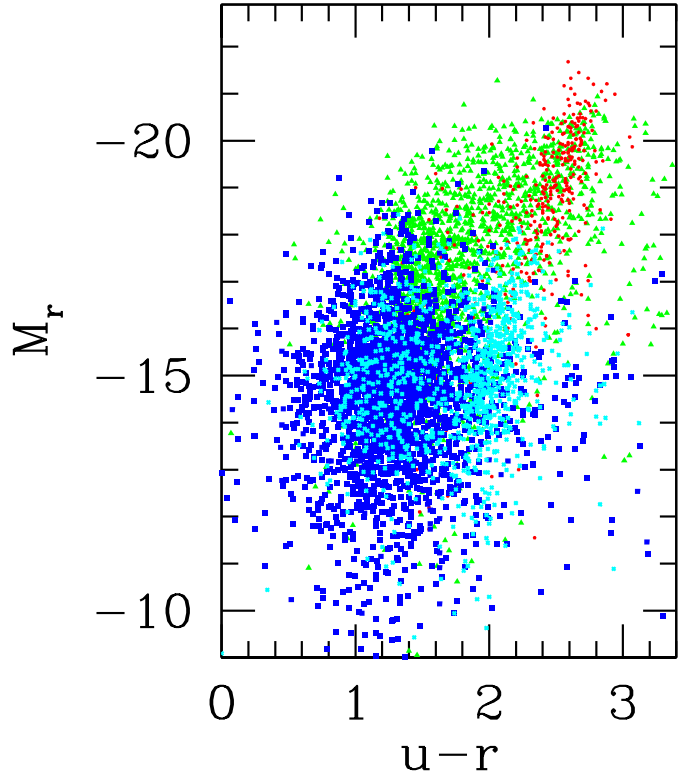}
\caption{Color-magnitude diagram of the local galaxies. Filled
triangles, filled rectangules, and croses represent spirals,
irregulars, and dwarf elliptical-like galaxies, respectively, whereas
soild dots designate ellipticals and lenticulars.
\label{fig14}
}
\end{figure}
%%%%%%%%%%%%%%%%%%%%%%%%%%%%%%%%%%%%%%%%%%%%%%%%%%%%%%%%%%%%%%%%%%%%%

\section{Properties of Local Galaxies}
\subsection{Luminosity and Color}

The absolute $r$-band magnitude and $u-r$ color are basic properties of 
a galaxy. The former is related directly to the mass of a galaxy, which is the
most important parameter in the formation and evolution of the galaxy, whereas
the latter is closely related to the star formation history. Because 
most of the sample galaxies do not have redshift-independent distances, we use
redshift-dependent distances for all of them to calculate $M_{r}$. 
We corrected the
motion of the LG following \citep{mou00} but the K-correction and evolution
correction are not applied in the derivation of $M_{r}$ because they are
negligibly small for local galaxies.

Figure~14 shows the $u-r$ color versus $M_{r}$ diagram of the 5836 galaxies
listed in the present morphological catalog. Some features should be
noted. First, galaxies with a different morphology occupy different
parts of the color-magnitude diagram. The so called red sequence is well
defined from $M_{r}\approx-13$ to $M_{r}\approx-22$ within which the 
luminosity-color relationship is well established.
The upper part of the red sequence consists of
elliptical galaxies and S0 galaxies (red color), whereas the lower part 
consists of dE-like galaxies (cyan color). They overlap around
$M_{r} \approx -17$, which suggests that $M_{r}=-17$ is the  bright limit 
of dwarf galaxies. Second, irregular galaxies (blue color) display
the  blue cloud. They are clustered around the mean values
of $M_{r} = -14.5 \pm3.3$ and $u-r = 1.36 \pm0.42$, respectively.
Some dE-like galaxies are located in
the blue cloud. They are mostly dE$_{blue}$ galaxies.
Third, spiral galaxies are located mostly in the green valley. The majority
of spiral galaxies are brighter than the dwarf limit of $M_{r}=-17$ but a
non-negligible fraction of spiral galaxies are fainter than $M_{r}=-17$.
This is due mainly to the late-type spiral galaxies of the Hubble stages
d, dm, and m. Finally, there are a number of galaxies whose photometry needs
to be refined. For example, galaxies located in the far red envelope of the red
sequence are probably due to poor photometry, even though some are due to
heavy extinction. In addition, $M_{r}$ of galaxies fainter than $M_{r}=-10$ 
are also due mostly to poor photometry.

%%%%%%%%%%%%%% Fig~15 FMrT.ps  %%%%%%%%%%%%%%%%%%%
\begin{figure}
\epsscale{.80}
\plotone{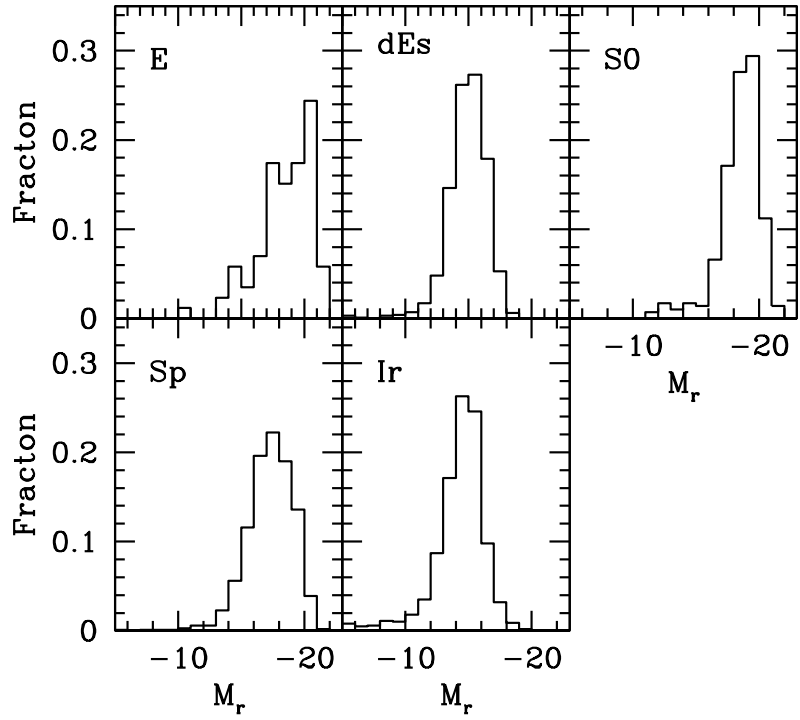}
\caption{Fractional frequency distributions of $M_{r}$ of 
the local galaxies. The 'dEs' represents the dwarf elliptical-like galaxies.
\label{fig15}
}
\end{figure}
%%%%%%%%%%%%%%%%%%%%%%%%%%%%%%%%%%%%%%%%%%%%%%%%%%%%%%%%%%%%%%%%%%%%%

\clearpage

Figure 15 shows the fractional frequency distributions indicating the 
luminosity distribution of local galaxies. The galaxies are grouped according
to their broad morphological types: E, dEs, S0, Sp, and Ir.
Although HII region-like BCD galaxies are supposed
to be dwarf galaxies similar to dE$_{blue}$, we grouped all the BCDs as a 
subtype of the Ir class because the majority of BCD components are imbeded in 
the irregular galaxies.  
Luminosity distributions of galaxies with different morphological types
differ significantly. Elliptical galaxies show a highly
asymmetric distribution that peaks at M$_{r}=-20.5$, whereas dEs and 
irregular galaxies show symmetrical distributions with similar
peak magnitudes of M$_{r}\approx-15$. Lenticular galaxies show a narrower
distribution than elliptical galaxies with a peak at M$_{r}=-19.5$. Among 
the three giant galaxies, E, S0, and Sp, spiral galaxies show the most 
symmetrical distribution with a peak luminosity at M$_{r}=-17.5$. Therefore, 
spiral galaxies are likely to be faintest among the three types of giant
galaxies. It is due to the dominance of late-type spirals which becomes
fainter toward the later Hubble stages \citep{ber60, str80, but07}.

Figure 16 presents the $u-r$ color distributions of the local galaxies which
are sorted according to the broad morphological types.
The color distributions depend
more strongly on the morphological types than the luminosity distribution.
The dependence of the color distribution on the morphological type is expected
because the color of a galaxy is a widely used proxy of the galaxy morphology. 
As shown in Figure 16, elliptical galaxies show the narrowest color distribution
centered on $u-r\approx2.5$. Approximately $90\%$ of elliptical galaxies
have $u-r$ colors between 2.2 and 3.0. The blue elliptical galaxies, which 
comprise $\sim9\%$ of all elliptical galaxies, have a blue tail.
Lenticular galaxies are also likely to have red colors similar to elliptical
galaxies. Approximately $73\%$ of lenticulars have $u-r$ colors 
redder than 2.2. On the other hand, 
spiral galaxies are likely to have blue colors. Approximately $78\%$ of 
spiral galaxies have $u-r$ colors bluer than $u-r=2.2$. This makes their color
distribution significantly different from that of lenticular galaxies. This
shows that the use of the colors of a galaxy as a proxy for the galaxy 
morphology
will make a significant contamination inevitable. The color distribution 
of dE-like galaxies shows a double peak, one at $u-r=1.2$ 
and the other at $u-r=1.9$. The blue peak is caused by the blue colors of
dE$_{blue}$ galaxies and the red peak is due to dE and dSph galaxies.
The irregular galaxies show a completely different color distribution from the
others. Their peak $u-r$ color is 1.2, which is 0.4 mag bluer than that of
spiral galaxies. Only $\sim5\%$ of irregular galaxies are redder
than $u-r=2.2$.

%%%%%%%%%%%%%% Fig~16 FurT.ps  %%%%%%%%%%%%%%%%%%%
\begin{figure}
\epsscale{.80}
\plotone{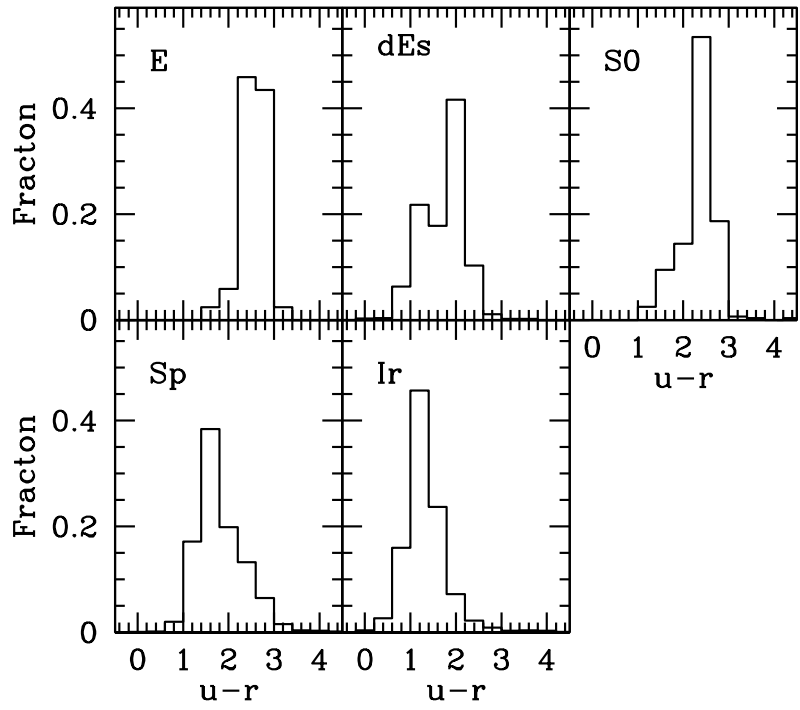}
\caption{Fractional frequency distributions of $u-r$ colors of the
local galaxies.
\label{fig16}
}
\end{figure}
%%%%%%%%%%%%%%%%%%%%%%%%%%%%%%%%%%%%%%%%%%%%%%%%%%%%%%%%%%%%%%%%%%%%%
\clearpage

%%%%%%%%%%%%%% Fig~17 glf.ps  %%%%%%%%%%%%%%%%%%%
\begin{figure}
\epsscale{.80}
\plotone{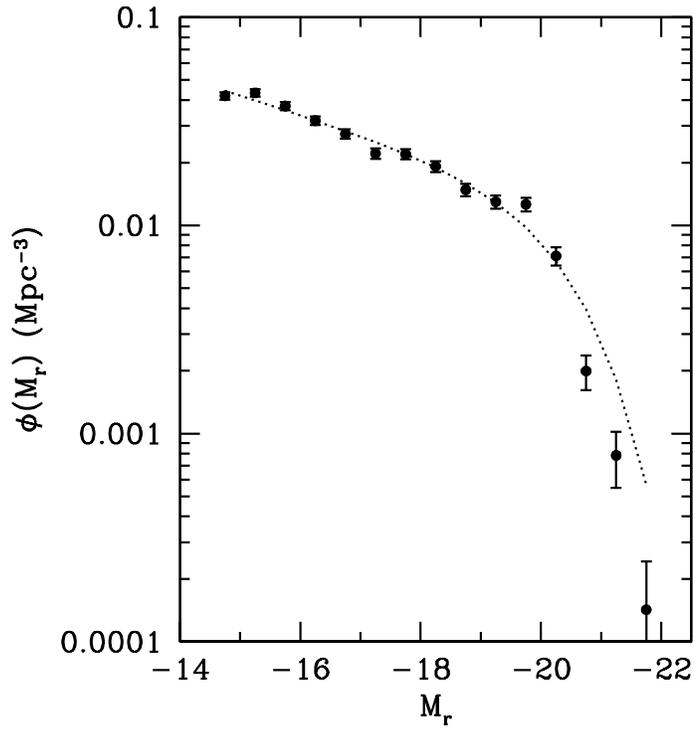}
\caption{Luminosity function of the
local galaxies. The best fit-Schechter function is plotted by the dotted line.
Errors are Poisson errors. 
\label{fig17}
}
\end{figure}
%%%%%%%%%%%%%%%%%%%%%%%%%%%%%%%%%%%%%%%%%%%%%%%%%%%%%%%%%%%%%%%%%%%%%
\clearpage

Figure 17 shows the luminosity function of the local galaxies, which extends to
$M_{r}\approx-15$, together with the best-fitted Schechter luminosity 
function \citep{sch76} of the form
\begin{equation}
\phi(M)dM=0.4~ln10~\phi^{\star}[10^{-0.4(M-M^{\star})}]^{\alpha+1}exp[-10^{-0.4(M-M^{\star})}]dM,
\end{equation}
where $M^{\star}$ is the characteristic luminosity of a galaxy 
and $\phi^{\star}$ is the number density in units of Mpc$^{-3}$.
We determined the three parameters, $\phi^{*}=0.015$, 
$\alpha=-1.22$, and $M_{r}^{*}=-20.58$ by minimizing $\chi^{2}$.
The drop at $M_{r} > -15$ is attributed to the observation
limit of the SDSS spectroscopic observation, which is set to $r_{pet}=17.77$.
This confines the volume-limited sample to galaxies brighter than 
$M_{r}=-15.2$. The rapid drop at $M_{r} < -20$ is somewhat different from the
luminosity functions of galaxies derived from the different volume-limited 
samples of different limiting magnitudes. \citet{cho07} derived the luminosity 
functions for six volume-limited samples from SDSS DR4plus. Their slopes
are much shallower than the slope of the present luminosity function. The
reason for the steeper gradient, i.e., the smaller $\alpha$ in the Schechter 
functions, for the present sample is that the local volume is less dense than
the regions outside the local universe. A similar trend for the steeper 
gradient in the luminosity functions of nearer samples was also observed 
among the volume-limited samples \citep{cho07}.

\subsection{Isophotal Radius}

The size of a galaxy can be defined in a variety of ways. We adopted the 
isophotal semi-major axis length as a representative radius of the local 
galaxies. Isophotal diameters can be affected by galactic extinction 
and inclination. We suppose that galactic extinction is negligible for the 
present sample due to the SDSS-covered area, however, inclination might affect
the isophotal diameter. There have been a number of studies on the effect of
inclination on the isophotal diameter \citep{hol58,dev76,pep86,jon96}. 
While \citet{hol58} assumed that the change in diameter due to 
the inclination of a galaxy is negligible, \citet{dev76} thought that
the inclination effects are need to be corrected.
However, as shown by \citet{pep86} and \citet{jon96}, it depends on
the optical nature of galaxies. In the case of spiral galaxies, their disks are 
believed to be optically thick \citep{dis89}, whereas the disks of lenticular
galaxies are thought to be optically thin \citep{mic93}. The isophotal 
diameters of optically thin disks increase with inclination.
The isophotal radius is calculated from the isophotal semi-major 
axis of the $r$-band image obtained from SDSS DR7. The isophotal semi-major and 
semi-minor axes are measured at a surface brightness ($\mu$) of 25
mag arcsec$^{-2}$ at five pass-bands ($u, g, r, i, z$) in the SDSS, 
and are given in units of pixels. The isophotal axis lengths in the pixels
are converted to kiloparsecs using a pixel size of 0.396 arcsec and
the distances used to calculate
the absolute magnitude. This isophotal radius is close to the Holmberg radius
which is measured at $\mu_{B}=26.5$ mag arcsec$^{-2}$. 

Figure~18 shows the 
frequency distribution of the isophotal radii of local galaxies sorted
according to their broad morphological types: E, dEs, S0, Sp, and Ir. 
As shown in Figure 18, the distributions of the isophotal radii of galaxies 
depend strongly on the morphological types of the galaxies. As expected,  
dE-like galaxies show the narrowest distribution, which peaked 
at $R_{iso}\approx1.5$kpc, and are limited to $\sim5$kpc. The Ir class of 
galaxies shows a similar narrow distribution with the same peak but it extends
to $\sim8$kpc. This is because most Ir galaxies are dwarf galaxies
with sizes similar to dEs. Some galaxies
larger than $R_{iso}\sim5$kpc are late-type spiral-like galaxies or peculiar
galaxies, some of which are merging galaxies. The ranges of isophotal radii 
of the ellipticals and lenticulars are similar, even though there are large
fluctuations in the ellipticals due to the small number of galaxies. 
Interestingly, the most probable sizes of the lenticulars and spirals are
similar to each other ($\sim3.5$kpc) whereas the ranges of the distributions
are significantly different. The sizes of the largest spirals are about two
times larger than those of the lenticulars. The same is true for those 
elliptical
galaxies which show a similar size distribution to that of the lenticular
galaxies. The reason for the double size of the spiral galaxies is 
thought to be the late accretion of the intergalactic material \citep{lar76,
mat89,van04,naa06,com14},
which makes the inside-out growth of galaxies similar to that assumed 
for the formation of the Milky Way.

%%%%%%%%%%%%%% Fig~18 isoA.ps  %%%%%%%%%%%%%%%%%%%
\begin{figure}
\epsscale{.80}
\plotone{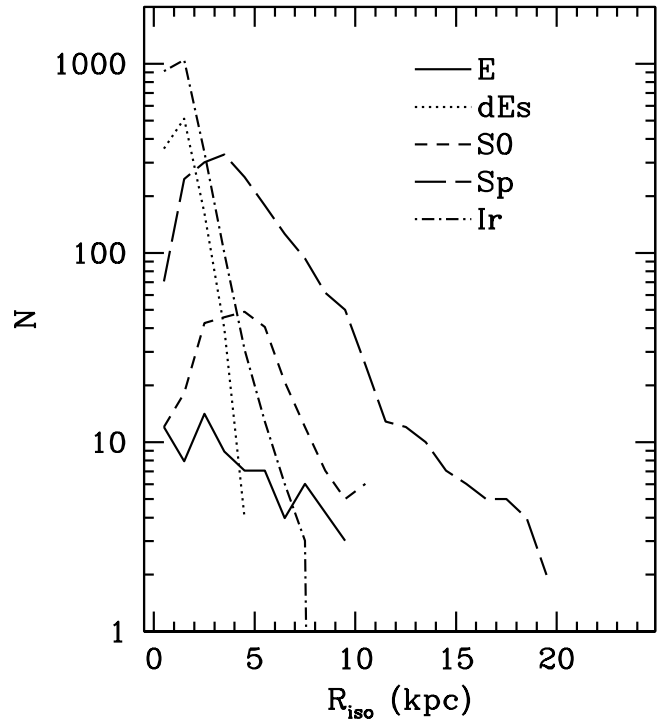}
\caption{Frequency distribution of isophotal radius ($R_{iso}$) of the
local galaxies.
\label{fig18}
}
\end{figure}
%%%%%%%%%%%%%%%%%%%%%%%%%%%%%%%%%%%%%%%%%%%%%%%%%%%%%%%%%%%%%%%%%%%%%
\clearpage

The basic properties of a galaxy such as mass, luminosity, radius, and 
morphology are believed to be determined by the galaxy formation process.
These parameters are supposed to be correlated with each other because they
have the same origin. 
The radius-luminosity relation \citep{she03,mci05} is an example of such a 
correlation. Because the galaxy size depends on the morphology, as shown
in Figure 18, the interplay between the size and luminosity is supposed to be
more complex. There have been many studies to investigate the 
morphology dependence of the radius-luminosity relation of galaxies since the
pioneering work by \citet{hub26}. Most of the previous studies used
proxies of galaxy morphology rather than morphology itself \citep{str01,
she03,mci05,ber14}. For example, \citet{she03} used the concentration index
as a proxy of the morphology. However, as is well known from previous studies
\citep{abr94,abr03,con03,kel04,lot04,par05,sca07,hue08,che11,cib13,wil14}, 
proxies are at best a crude approximation of galaxy morphological types.  
In this regard, \citet{guo09} distinguished early-type galaxies and late-type
galaxies by visual inspection of the images for an analysis of the structure
of central galaxies in groups and clusters.
Therefore, the morphological types presented in this study will be
useful to explore the interplay between the size and luminosity of galaxies.
We are planning to do this in a series of forthcoimng papers. 
We are particularly
interested in the radius-luminosity relation of dwarf galaxies.

%%%%%%%%%%%%%% Fig~18 > Fig19  Fb2a.ps  %%%%%%%%%%%%%%%%%%%
\begin{figure}
\epsscale{.80}
\plotone{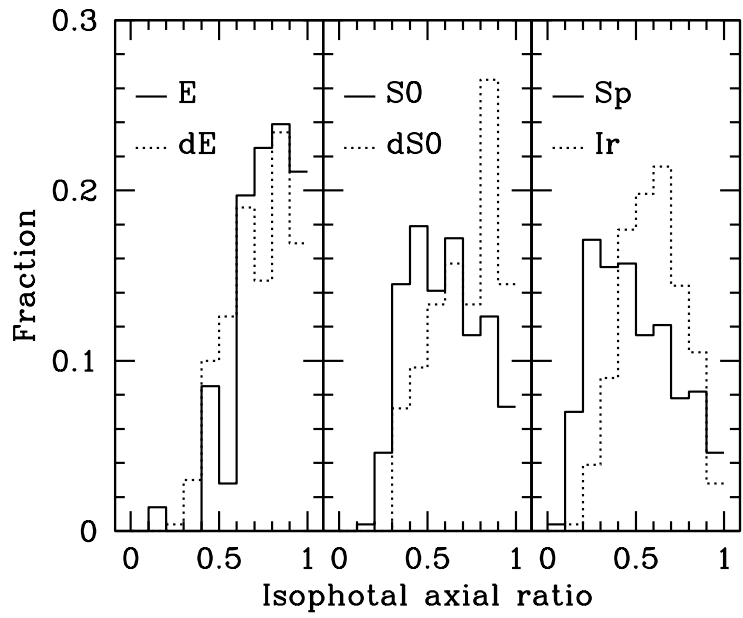}
\caption{Fractional distribution of isophotal axis ratios of the local
galaxies. E and dE (dwarf elliptical-like galaxies except dwarf lenticulars) 
are presented in the left panel, S0 and dS0 in the middle panel,
and Sp and Ir in the right panel. 
\label{fig19}
}
\end{figure}
%%%%%%%%%%%%%%%%%%%%%%%%%%%%%%%%%%%%%%%%%%%%%%%%%%%%%%%%%%%%%%%%%%%%%
\clearpage

\subsection{Axis Ratio}     % IsoA

The axial ratio of a galaxy ($b/a$) is closely related to the intrinsic shape
of the galaxy, which is determined by its dynamical structure. Figure 19 shows
the fractional frequency distributions of the axis ratios of 5836 galaxies 
sorted according to their morphological types.
The isophotal semi-major and semi-minor axes in $r$-band, which
are measured at a surface brightness of 25 mag arcsec$^{-2}$, are used.
Figure 19 plots the three morphological types of giant
galaxies (E, S0, and Sp) compared to their dwarf cousins (dE and dS0) for E 
and S0 galaxies and irregular galaxies for spiral galaxies. 
The axis ratio distributions of the ellipticals and
dE (left panel) are similar, whereas a 
significant difference was observed in the lenticular and dwarf lenticular
galaxies (middle panel) as well as in spiral galaxies (right panel). The solid
lines represent the giant galaxies and the dotted lines indicate the dwarf
counterparts. 

Although the general shapes of the axis ratio distributions of the giant
elliptical galaxies and dE galaxies are
similar, there are some differences between the two. Dwarf
elliptical galaxies have a higher fraction of galaxies with $b/a < 0.5$ than
giant elliptical galaxies, even though the smallest $b/a$ is observed 
in a giant elliptical galaxy. 
On the other hand, the morphological type of this galaxy (NGC~2768) is
controversial. It is classified as E6 in RC3, whereas the morphological
type adopted by the NED is S0, which was taken from The Carnegie Atlas of 
Galaxies \citep{san94}. The more flattened structure of dE
galaxies is consistent with those observations showing fast rotation in a 
considerable fraction of dE galaxies \citep{geh10}. On the
contrary, the general similarity of the axis ratio distributions of the
giant and dwarf elliptical galaxies suggests that a large fraction of dwarf
elliptical galaxies are dispersion-supported systems like giant elliptical
galaxies. 

The dissimilarity of the axis ratio distributions of giant and dwarf
lenticular galaxies is quite interesting because dS0s are
considered to be low-luminosity versions of S0s \citep{agu05}.
The axis ratio distribution for dS0s in Figure 19 seems to show 
well that the dS0s should not be thought of as small versions of S0s.
Rather, they are close to dE galaxies because their axis ratio 
distribution is similar to that of dE galaxies. The lack of dS0s
with an isophotal axis ratio ($b/a$) less than 0.3 suggests that
they are intrinsically thicker than S0s. The lack of small axis ratios is 
also observed in dE and irregulars, suggesting less flattened
shapes for these galaxies. The thick intrinsic shape of irregular 
galaxies is consistent with earlier studies \citep{hei72,ber88,roy10}.

The dissimilar distributions of the axis ratios of spiral 
galaxies and irregular galaxies 
shown in the right panel of Figure 19 are believed to 
be caused by the real difference in their dynamical structures because
the statistical noises are small enough to be ignored due to the large number 
of spiral and irregular galaxies. In the case of lenticular galaxies, the axis
ratio distribution of which is supposed to be similar to that of the spiral
galaxies. We can see a large fluctuation in the histogram due to the small
number of lenticular galaxies, which is about one-seventh that of the spiral 
galaxies. The larger fraction of small axis ratios in spiral galaxies is due
to the rotating disks of spiral galaxies, the orientations of which are random, 
whereas the lack of small axis ratios in the irregular galaxies is believed to
be due to the triaxial nature of irregular galaxies \citep{ber88}. 

%%%%%%%%%%%%%% Fig~19 > 20 Fb2adE.ps  %%%%%%%%%%%%%%%%%%%
\begin{figure}
\epsscale{.80}
\plotone{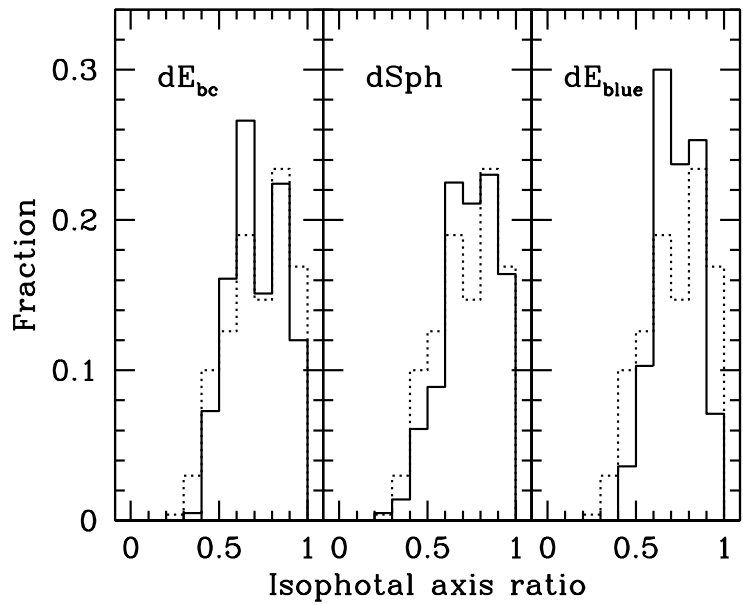}
\caption{Fractional distribution of isophotal axis ratios of the local
galaxies. Blue-cored dwarf ellipticals (dE$_{bc}$) in the left panel, dwarf
spheroidals (dSph) in the middle panel, and blue dwarf ellipticals (dE$_{blue}$)
in the right panel. Histograms plotted by the dotted lines are those for
the dwarf ellipticals (dE). 
\label{fig20}
}
\end{figure}
%%%%%%%%%%%%%%%%%%%%%%%%%%%%%%%%%%%%%%%%%%%%%%%%%%%%%%%%%%%%%%%%%%%%%
\clearpage

Figure~20 shows the fractional distribution of the axis ratios of the dwarf 
elliptical-like galaxies; dE$_{bc}$ galaxies in the left panel, 
dSph galaxies in the middle panel, and dE$_{blue}$ galaxies in the right panel.
The axis ratio distribution of dE galaxies (dotted lines) is also 
plotted for comparison. Due to the small numbers, there are
large fluctuations in the histograms but there seem to be some differences
among the subtypes of dE-like galaxies. The most pronounced
difference is observed between dE and dE$_{blue}$ galaxies. It is due to
the narrow range of axis ratio distribution of dE$_{blue}$ galaxies with a
lower limit of $b/a=0.4$. Approximately 80$\%$ of  dE$_{blue}$ galaxies
have axis ratios between $b/a=0.6$ and $b/a=0.9$. This siggests that they 
are mostly dispersion-supported systems but are not as round as dSph galaxies
and dispersion-supported dE galaxies.
The lack of a systematic difference between dE and dE$_{bc}$ is consistent
with the morphological similarity between the two types
of galaxies, except for the presence of blue cores in dE$_{bc}$ galaxies.  
The axis ratio distribution of dSph galaxies is similar to that of dE galaxies
but the lower fraction of galaxies with small $b/a$ appears to be
real because they are systematically lower than that of dE galaxies for a small
axis ratio ($b/a \lesssim 0.4$). The lack of galaxies with a small $b/a$ is 
consistent with the kinematic observations of dSph galaxies \citep{sal12}.

\subsection{Morphological Properties of Dwarf galaxies}
\subsubsection{Dwarf Ellipticals and Blue-cored Dwarf Ellipticals}

The morphology of dwarf elliptical galaxies is characterized by their
round appearance and red colors, similar to that of giant elliptical
galaxies. On the other hand, there are some differences in the photometric
properties between dE and giant ellipticals. For example, dE
galaxies exhibit lower surface brightness and a shallower
luminosity gradient than giant elliptical galaxies.
The mean luminosity of dE galaxies is $M_{r} = -15.9 \pm 1.0$,
which is approximately 3.5 mag fainter than the mean luminosity ($M_{r}
\approx -19.5$) of elliptical galaxies listed in the present catalog. 
\citet{wir84}, \citet{kor85}, and \citet{kor09} argued that these 
differences result from the 
structural difference between dE and giant ellipticals. 
Therefore, dE galaxies cannot simply be a small version of
giant elliptical galaxies (Kormendy \& Bender (2012, and references therein).
The real dwarf
counterparts of giant elliptical galaxies are compact elliptical galaxies, 
exemplified by M32 \citep{fer94,kor09}. They have sizes comparable to dwarf 
elliptical galaxies but have a high surface brightness. On the other hand,
\citet{jer97} and \citet{gra03} agrued that the scaling relations of the dwarf
elliptical galaxies are continuations of those of giant elliptical galaxies
because the nonlinear relations between the absolute magnitude and the 
effective surface brightness or effective radius are consequences of the two
linear relations between the stellar concentration, quantified by the Sersic 
index $n$, and stellar mass or central surface brightness. Their structural
characteristics are not distinguished from those of giant elliptical 
galaxies (Graham 2013, and references therein).
Whether or not dE galaxies are distinct populations, 
they are thought to be a mixture of kinematically heterogeneous
populations \citep{geh10}, the origins of which are thought to be
different \citep{lis09,tol12,lis13}.

The three subtypes of dE-like galaxies, dE$_{un}$, dE$_{n}$
and dE$_{bc}$, have similar structure except for the nuclear morphology. 
Normally, the blue cores of dE$_{bc}$ galaxies are larger than the star-like
nuclei in dE$_{n}$ galaxies but in some cases, they appear small.
However, we classify them as dE$_{bc}$ galaxy if 
their colors are blue enough to assume the presence of young stellar 
populations. The number of dE$_{bc}$ galaxies that have a small core is
negligibly small. The global colors of dE$_{bc}$ galaxies are similar to
those of dE$_{un}$ and dE$_{n}$ galaxies. However, in some cases, their colors
appear quite blue close to dE$_{blue}$ galaxies. 

The well-known examples of dE$_{bc}$ galaxies are NGC 185 \citep{hod63}
and NGC 205 \citep{hod73}, both of which are satellite galaxies of M 31. 
The blue cores of dE$_{bc}$ galaxies are thought to be due to recent star
formation. Different scenarios have been proposed for the origin of the gas in 
the center of dE$_{bc}$ galaxies. The gas left in the spiral and irregular 
galaxies which are transformed into dE by ram pressure stripping 
\citep{gun72} or galaxy harassment \citep{moo96,moo98} is widely considered for
the origin of the central star formation of dE$_{bc}$ galaxies in cluster
environment \citep{lis06,lis07,pak14}. On the other hand, gas accretion from
the cosmic web \citep{hal12} can be resposible for the isolated dE$_{bc}$ 
galaxies such as IC 125 \citep{guq06}. 
In addition, it is possible that small cores such as star-like blue nuclei are
made from recycled gas from evolved stars \citep{bos08,hal12}. 

\subsubsection{Dwarf Spheroidals}

The morphology of dSph galaxies is similar to that of dwarf 
elliptical galaxies.
They share a round appearance and red color, lower surface brightness, and
shallower gradient in the surface brightness than elliptical galaxies. 
This is why many studies did not distinguish them \citep{kor12}.
On the other hand, there are
some differences in their morphological properties. On average, they are 
fainter than dE galaxies. They have a mean 
luminosity of $M_{r} = -14.4 \pm 1.2$, which is $\sim1.5$ mag fainter than the
dE galaxies. In addition, they differ in their surface brightness 
distribution. The surface brightness of dSph galaxies is, on 
average, fainter than that of dE galaxies. Also, dSph 
galaxies show a gentler gradient in their surface brightness
distribution than dE galaxies. The low luminosity and low 
surface brightness of dSph galaxies are consistent with the relation
between the luminosity and surface brightness of a galaxy \citep{san84}.

In addition to the difference in surface brightness between dSph
galaxies and dE galaxies, there are some properties that are
quite different. The presence of a blue core is barely observed in the dSph
galaxies, whereas a significant fraction of dE
galaxies have blue cores. Some difference in the axis ratio 
distributions is present, showing smaller fractions of highly flattened dSph
galaxies. Because axis ratios, i.e., ellipticities are 
closely related to the degree of rotational support \citep{bin78},
dSph galaxies are believed to be mostly a dispersion-supported
system, whereas dE galaxies are considered to be partially
supported by rotation. This picture of kinematic properties of dSph
galaxies is frequently addressed in previous observations \citep{wal09,tol12}.
On the other hand, the distinction of ellipticity distributions between dSph
and dE is not as clear as that in the fast and slow 
rotating elliptical galaxies observed in the ATLAS \citep{wei14} due to the
fact that the fractions of rotationally supported dEs and 
dispersion-supported dEs are similar \citep{geh10}. However, the
differences in photometry as well as in kinematics between the dSph and dE
galaxies suggest that they have different origins. It is quite likely that
dSph galaxies are primodial objects formed at $z > 7$ \citep{kor14}, while dE
galaxies, at least fast rotating dE galaxies, are stripped late type galaxies.

\subsubsection{Blue Dwarf Ellipticals}

The morphological distinction of dE$_{blue}$ galaxies from other 
dwarf galaxies is one of the most successful uses of color images in morphology
classification. While the blue colors of blue elliptical
galaxies \citep{str01}, which distinguish them from normal elliptical
galaxies, are believed to be due to recent star formation in the core 
regions, the colors of the dE$_{blue}$ galaxies are due to 
global star formation that makes their colors much bluer than the dwarf
elliptical galaxies. These galaxies have similar colors to those of irregular 
galaxies. Their shapes, however, are quite similar to those of dE 
galaxies although their axis ratio distribution is somewhat different.
dE$_{bc}$ are
considered to be much different from the dE$_{blue}$ galaxies
because dE$_{bc}$ galaxies have red colors outside the blue core. 
In some aspects, the dE$_{blue}$ galaxies are
similar to HII region-like BCDs or dI. The former shows somewhat
bluer colors than dE$_{blue}$ but with a similar shape, and the latter shows
a lack of axial symmetry compared to dE$_{blue}$. 

\subsubsection{Dwarf Lenticulars}

The morphology of dwarf lenticular galaxies (dS0) is similar to their
giant cousin, lenticular galaxies (S0), in the sense that they show subtle 
structure, such as the bulge and disks in S0s. On the
other hand, they are significantly different from S0s in many aspects. 
The morphology of the central excess of light in dwarf lenticular galaxies 
appears to be lens-like \citep{san84} and much less massive than a bulge. 
In some cases, the lens component is quite large in size but their
luminosity seems to be low. They are likely to have star-like nuclei, 
which causes them to be classified as dS0$_{n}$, similar to the nucleated dwarf
elliptical galaxies, dE$_{n}$. Their colors are also similar to dwarf 
ellipticals. In addition, it is quite rare to
have bars inside and there appears to be no inner and outer
rings in the dwarf lenticular galaxies. The absence of 
inner and outer rings in dwarf lenticular galaxies is expected if
they are stripped, former late-type galaxies. Rings are a phenomenon mainly of
massive early-type galaxies. They are virtually never seen among
extreme late-type spirals or magellanic irregulars \citep{but96,but98}.
On the other hand, they are likely to show
disrupted features in the outer parts of the galaxies. These features are
probably remnants of spiral arms. If this is the case, dwarf spirals can
be formed in disks as small as those of dwarf lenticular galaxies but can not
survive long due to the lack of sufficient mass to protect them from  
tidal or hydrodynamic disturbances.

\subsubsection{Blue Compact Dwarfs}

Compact galaxies are distinguished from others by their high surface 
brightnesses which are brighter than 20 mag arcsec$^{-2}$ \citep{zwi71}.
A BCD galaxy has very blue color indicative of a likely
starburst. BCDs
show a range of morphologies in the present catalog. One extreme case
is a BCD whose morphology is characterized by a HII region-like appearance,
and the other extreme case is characterized by localized compact regions of
intense bursts of star formation within a larger irregular galaxy. Examples of
the two extreme cases are presented in the two rightmost columns of the
bottom row in Figure 5. 

The HII region-like BCDs similar to those observed by \citet{thu81} who chose 
their samples mostly from the earlier objective prism surveys.
%% \citep{har56,mar67,mar69a,mar69b,mar71,mar72,mar73,mar74}.
The shape of these galaxies is
similar to the small-sized dE$_{blue}$ galaxies but they have deep bluer
colors. Because there is no clear color distinction between these two types 
of galaxies, they can be classified as genuine BCDs or isolated extragalactic  
HII regions \citep{sar70}. On the other hand, the selection of small-sized 
dE$_{blue}$ galaxies from the dE$_{blue}$ sample of galaxies is also subjective.
The other types of BCDs are basically irregular galaxies with localized 
compact regions of intense star-bursts. The size of the compact regions of 
extreme star-burst varies from the size of HII regions to the galaxy size. 
Therefore, it is also subjective to set a size-limit when selecting BCDs. 
In the present study, a galaxy is classified as BCD if the size of the 
compact regions of extreme star burst is larger than $\sim20\%$ of the galaxy
size. In order to distinguish the irregular galaxies with small star burst 
regions, we assign Im/BCD as the morphological type of these galaxies.
They are amount to $\sim20\%$ of irregular galaxies.

\subsubsection{Dwarf Irregulars}

The majority of galaxies in the local universe are dwarf galaxies
if we consider irregular galaxies fainter than M$_{r}\approx -17$
to be dwarf galaxies, even though we classify them as Im galaxies because of
morphologies resembling the late-type spirals. dI
galaxies show a range of morphologies from very simple shapes to a high degree
of complexity. dI with a simple morphology resemble dE$_{blue}$ 
or BCDs. The only difference is the degree of roundness. 
dI galaxies with complex morphologies are caused mainly by sites
of intense star formation. They are too small to drive the global density waves.
Star formation in dI galaxies is believed to be caused by 
self-propagating star formation \citep{ger78}. Interactions with  
neighboring  galaxies makes their morphology rounder and rounder 
if the interactions occur frequently.
The morphology of some dI galaxies show a transition type
from dI to dE/dSph or vice versa.

%%%%%%%%%%%%%% Fig~21  FdenT.ps  %%%%%%%%%%%%%%%%%%%
\begin{figure}
\epsscale{.80}
\plotone{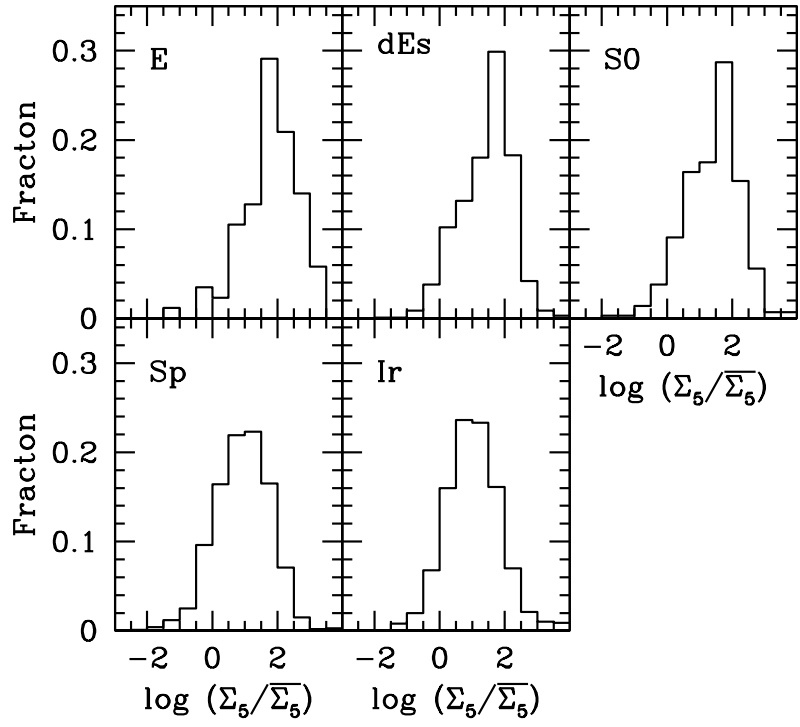}
\caption{Fractional distributions of the local background density 
$\Sigma_{5}$ normalized by the mean local background 
density $\bar{\Sigma_{5}}$.
\label{fig21}
}
\end{figure}
%%%%%%%%%%%%%%%%%%%%%%%%%%%%%%%%%%%%%%%%%%%%%%%%%%%%%%%%%%%%%%%%%%%%%
\clearpage

\section{Environment Dependence of Morphology}
\subsection{Local Background Density}

The morphology of a galaxy is dependent on its environment. This dependence
has been represented by the morphology-density relation since the pioneering 
work of \citet{dre80}. There are a variety of ways to measure the local 
background density of a galaxy \citep{mul12}. Recently, \citet{ann14} explored
the dependence of disk morphology on the local background density
derived from the $n$th nearest neighbor method with $n=5$, $\Sigma_{5}$,
which is defined as
\begin{equation}
\Sigma_{5}= {5 \over {4\pi {r_{p,5}}^{2}}},
\end{equation}
\noindent{where $r_{p,5}$ is the projected distance to the fifth nearest
galaxy brighter than $M_{r}=-15.2$ with $|\Delta V| < 1000$kms$^{-1}$.} 
We normalized $\Sigma_{5}$ by the mean local background 
density ($\bar{\Sigma_{5}}$). 

Figure~21 shows the fractional density distribution of local galaxies sorted by
the broad morphological types.  As shown in Figure 21, the distributions of
the local background density of galaxies with different morphology differ
greatly. Elliptical
galaxies are most frequent in the high-density regions, forming a narrow peak 
at the high-density regions with an extended tail toward the low-density 
regions, whereas spiral and irregular galaxies show approximately 
symmetrical distributions with a 
peak density at log ($\Sigma_{5}/\bar{\Sigma_{5}})=0.75$. Lenticular
and dE-like galaxies show similar density distributions to
that of elliptical galaxies with a slightly larger fraction of galaxies in 
the low-density regions than elliptical galaxies. It is worth noting 
that dE-like galaxies do not obey the luminosity-density 
relation \citep{par07} because it dictates that dwarf galaxies which are
basically much less luminous than elliptical and lenticular galaxies are 
located preferentially in the low-density regions. This means 
that the morphology of a galaxy, particularly a dwarf galaxy, can be affected
greatly by other environmental properties. The interactions with large
galaxies to which most of dwarf galaxies are bound as satellite galaxies
are considered to be the main cause of the secondary parameter 
affecting the dwarf galaxy 
morphology. The distance to the nearest neighbor ($r_p$) appears to be a good
measure of the galaxy environment because the effect of interactions depends 
on the separation between the interacting galaxies.

%%%%%%%%%%%%%% fig21  > Fig~22  TFden.ps  %%%%%%%%%%%%%%%%%%%
\begin{figure}
\plotone{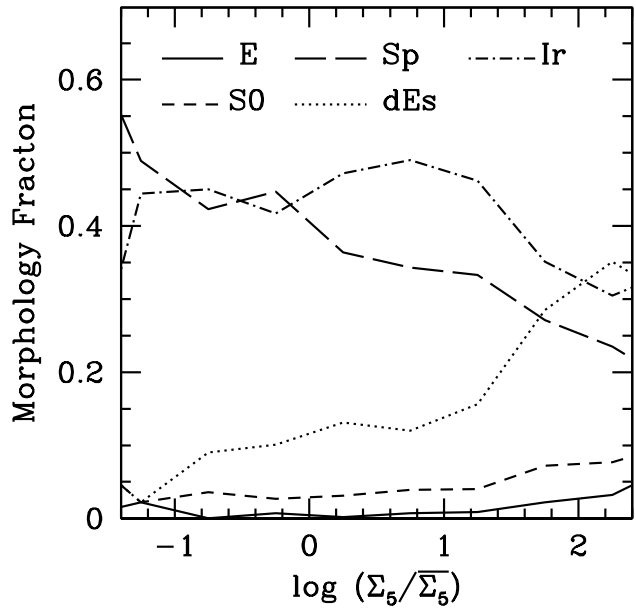}
\caption{Morphology fraction as a function of the local background density.
\label{fig22}
}
\end{figure}
%%%%%%%%%%%%%%%%%%%%%%%%%%%%%%%%%%%%%%%%%%%%%%%%%%%%%%%%%%%%%%%%%%%%%
\clearpage

To see the dependence of the galaxy morphology on the local background density,
the fraction of galaxy morphological types is plotted as a function of the
local background density (Figure 22).
In general, they follow the morphology-density relation, i.e.,
the fractions of early type galaxies (E, S0, and dEs) increase with 
increasing local background density, whereas those of late type 
galaxies (Sp and Ir) decrease with increasing density. Although the general
pattern of increasing fractions with increasing local background density is
the same for the three classes of early type galaxies, there are some 
differences between them. Fractions of the two early type giant 
galaxies (E and S0) show an almost flat
distributions at log ($\Sigma_{5}/\bar{\Sigma_{5}}) <1 $ and gradual increase
thereafter, whereas that of the early type dwarf galaxies (dEs) shows a
nearly monotonic increase with increasing density with a sudden change of
the slope at log ($\Sigma_{5}/\bar{\Sigma_{5}}) \approx 0.8$. 
Irregular galaxies show a nearly opposite trend from that of dEs
at log ($\Sigma_{5}/\bar{\Sigma_{5}}) \gtrsim 0.8$ and a somewhat flat   
distriobution at log ($\Sigma_{5}/\bar{\Sigma_{5}}) \lesssim 0.8$ with a
local minimum around log ($\Sigma_{5}/\bar{\Sigma_{5}})\approx-0.8$.
On the other hand, spiral galaxies demonstrate behavior opposite of the general
trend for early-type galaxies, that is, an almost monotonically decreasing 
spiral fraction with increasing local background density. Also, 
there is not much variation in the fractions of Hubble stages and bar types
of spiral galaxies along the local background density except for an increasing
early spiral fraction and a decreasing late spiral fraction with increasing 
local background density in high-density regions, 
log ($\Sigma_{5}/\bar{\Sigma_{5}}) >1$.

To observe the dwarf morphology dependence on the local background density, we
plot the distribution of the morphology fractions of dwarf 
galaxies as a function 
of the local background density in Figure 23. A clear difference in the 
morphology distribution can be seen between the two
groups of dwarf galaxies. One group of dwarf galaxies (dE, dSph, and dS0) shows 
similar fractions with an increasing trend for the density ranges of 
log ($\Sigma_{5}/\bar{\Sigma_{5}}) < 1.2$ and become different fractions at
the higher density regions. The other group (dE$_{bc}$ and dE$_{blue}$) shows
dominant fractions at local background densities less than the mean 
local background density. They comprise more than $90\%$ of the dEs galaxies
at log ($\Sigma_{5}/\bar{\Sigma_{5}}) < 0$, and decrease with increasing
density to become the same fractions as those of the former group at 
log ($\Sigma_{5}/\bar{\Sigma_{5}})\approx1.3$. The predominance of
dE$_{blue}$ galaxies in the low-density regions seems to reflect the 
environmental dependence of the star formation rate observed in the 
local universe, and higher star formation rate in the low-density
regions \citep{lew02,gom03,tan04}, in accord with down-sizing \citep{cow96}.
The high fraction of star forming dwarfs in the low-density regions is due to
the dependence of collapse time ($\tau$) of the pre-galactic cloud on the 
background density ($\tau \propto \rho^{-1/2}$). In the low-density regions,
a large amount of gas is left to make dwarf galaxies in the present epoch,
including dE$_{blue}$ galaxies.
The preference for dE$_{bc}$ galaxies in low-density regions can be understood
if late gas accretion is the main cause of the central star formation in the
isolaed dE$_{bc}$ galaxies. 

%%%%%%%%%%%%%% Fig~23 --> Fig~22   -> Fig~23 FdendE.ps  %%%%%%%%%%%%%%%%%%%
\begin{figure}
\plotone{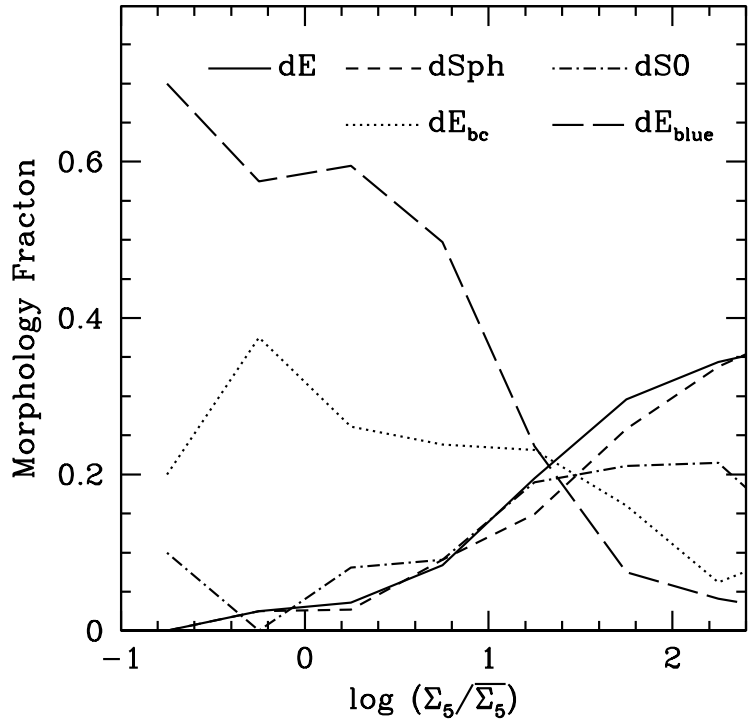}
\caption{Morphology fraction of dwarf elliptical-like galaxies as 
a function of the local background density.
\label{fig23}
}
\end{figure}
%%%%%%%%%%%%%%%%%%%%%%%%%%%%%%%%%%%%%%%%%%%%%%%%%%%%%%%%%%%%%%%%%%%%%
\clearpage

Although the local background density appears to be the primary parameter 
dictating
the morphology of a galaxy, its role is different for giant galaxies and 
dwarf galaxies. For giant galaxies, it constrains the galaxy morphology 
through the star formation efficiency, which depends on the dynamical 
timescales ($\tau \propto \rho^{-1/2}$) at the initial collapse phase and the
subsequent hierarchical merging phase of galaxy formation. In contrast, for
dwarf galaxies, it constrains the interaction rates with giant galaxies.
In cases of dwarf galaxies that are bound to giant galaxies as satellites,
the interaction rate is inversely proportional to the orbital period 
that is determined by the dynamical timescales.
Therefore, dE, dSph, and dwarf lenticulars,
which are likely to be located in the high-density regions, are
supposed to lose their gas quickly due to the frequent interactions with 
large galaxies. On the other hand, dE$_{blue}$ galaxies have fewer
interactions than other dE-like galaxies to retain their gas
longer. In the case of the dE$_{bc}$, the role of local
background density seems to be complicated. Since dE$_{bc}$
are similar to dE galaxies except for the blue core, their red 
color outside the nucleus suggests early removal of gas by frequent 
interactions with giant galaxies, but there is leftover gas in their 
environment to accrete the gas removed from dEs to make blue core.

\subsection{Nearest Neighbor}

The morphologies of galaxies are affected by the neighboring galaxies.
One of the measures of the influence of the neighbor galaxies is the projected
distance, $r_{p}$. This can be expressed in absolute units such as Mpc, but
$r_{p,n}$, which is normalized by the neighbor's virial radius,
is a good measure of the local environment because the morphology of a galaxy
is affected significantly by the neighbor galaxy when it is located within
the neighbor's virial radius \citep{par08,par09}. The virial radius of a 
galaxy is defined as the radius where the mean density inside the 
radius becomes the virialized density, which is set to
766 $\bar{\rho}$ \citep{par07}. The virial radius of a galaxy was calculated 
according to that described in \citet{par07}.
We derived $r_{p}$ of the 5836 local 
galaxies using the same neighbor search
conditions as those applied to derive the local background density.

%%%%%%%%%%%%%% Fig~24 --> Fig~23 -> Fig~24 Dvirmpc.ps  %%%%%%%%%%%%%%%%%%%
\begin{figure}
\plotone{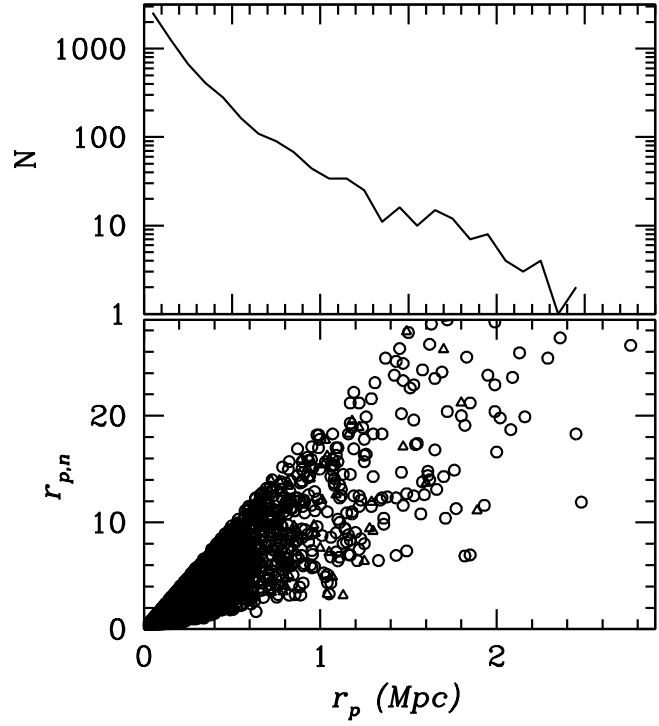}
\caption{Projected distance to the nearest neighbor galaxy $r_{p}$.
The frequency distribution of $r_{p}$ (Mpc) is plotted in the upper panel
and the scatter diagram of $r_{p,n}$ and $r_{p}$ (Mpc) is presented 
in the lower panel.
\label{fig24}
}
\end{figure}
%%%%%%%%%%%%%%%%%%%%%%%%%%%%%%%%%%%%%%%%%%%%%%%%%%%%%%%%%%%%%%%%%%%%%
\clearpage

Figure 24 shows the frequency distribution of $r_{p}$ (Mpc; upper
panel) and the scatter diagram of $r_{p,n}$ and $r_{p}$ (Mpc).
The frequency distribution in the upper panel shows a very rapid decrease
of number of galaxies with increasing $r_{p}$. Approximately $40\%$ of the 
local galaxies have a neighbor galaxy within $r_{p}=0.1$Mpc and similar
fractions of the local galaxies are located inside the virial radius of the 
nearest neighbor. Owing to the rapid decrease of the frequency distribution 
of $r_{p}$, the number of galaxies that have their nearest neighbors  
at $r_{p} > 1$Mpc is only 204 ($3.5\%$) and it is reduced to 66 ($\sim 1\%$)
for $r_{p} > 2$Mpc.
A good correlation is observed between $r_{p,n}$ and $r_{p}$ (Mpc) with
an upper envelope of an almost straight line. The scatter is caused by
the wide range of virial radii of galaxies. The upper envelope is due to 
the smallest dwarf galaxies while the lower envelope is due to the largest 
giant galaxies. Because of the small number of large giant galaxies, the lower
envelope is not so straight as the upper envelope.

%%%%%%%%%%%%%% Fig~25 --> Fig~24 > Fig25  FTsepvir.ps  %%%%%%%%%%%%%%%%%%%
\begin{figure}
\plotone{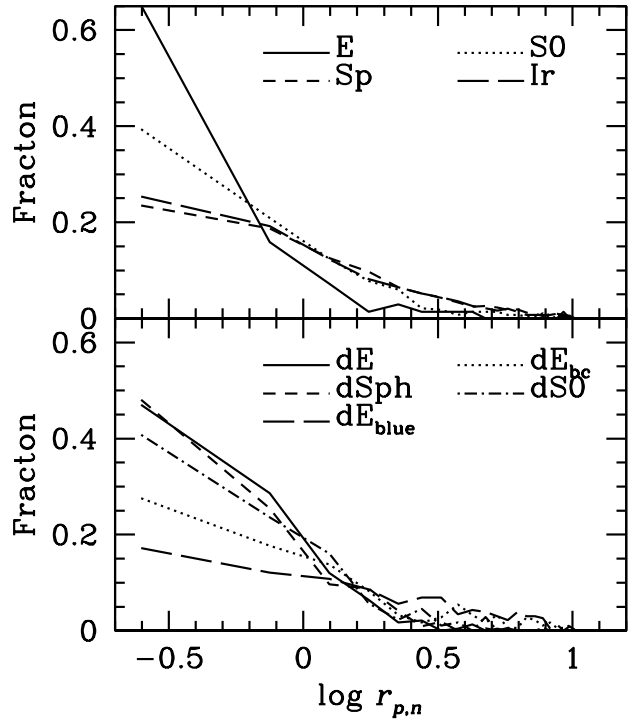}
\caption{Fractional distributions of morphological types as a function 
of $r_{p,n}$, four broad types of E, S0, Sp, and Ir in the upper panel,
and dEs in the lower panel.
\label{fig25}
}
\end{figure}
%%%%%%%%%%%%%%%%%%%%%%%%%%%%%%%%%%%%%%%%%%%%%%%%%%%%%%%%%%%%%%%%%%%%%
\clearpage

Figure 25 shows the fractional distributions of the morphological types as a 
function of $r_{p,n}$.
It is apparent that the majority of local galaxies are located
within the virial radius of the nearest neighbor. Among the morphological 
types of giant galaxies, the elliptical galaxies are
most likely located within the virial radius of the nearest neighbor, whereas
more than half of the spiral and irregular galaxies are located outside
the neighbor's virial radius. This means that the effects of hydrodynamical 
interactions with neighboring galaxies are more important in elliptical 
galaxies because hydrodynamical interactions are likely to be effective
when a galaxy is located within the neighbor's virial radius. Lenticular 
galaxies show an intermediate distribution between elliptical galaxies
and spiral galaxies. Among the dE-like galaxies, dE, dSph, and 
dS0 galaxies show fractional distributions similar to that of lenticular 
galaxies, whereas dE$_{blue}$ galaxies show a more extended tail toward large
$r_{p,n}$. More than $\sim40\%$ of dE, dSph, and dS0 galaxies are located 
at $r_{p,n} > 0.5$. The dE$_{bc}$ galaxies show intermediate distribution
between dE and dE$_{blue}$. An examination of the morphology fractions of
the 5 broad morphological types as a function of $r_{p,n}$ shows similar
trends to those observed in Figure 22 where the morphological fractions are
plotted as a function of the local background density.
That is, the three early types (E, S0, and dEs) show increasing 
fractions with decreasing $r_{p,n}$ and the two late types (Sp and Ir)
show the opposite trend. 

\clearpage

Figure 26 shows the early type fractions of the target galaxies as a function of
$r_{p,n}$. The early-type fractions of the target galaxies increase
with decreasing $r_{p,n}$. However, the increasing rates depend on
the neighbor's morphology. Here, early type represents three broad types
of galaxy, E, S0, and dEs, whereas late type represents spiral and
irregular galaxies. The early type fractions increase rapidly with decreasing
$r_{p,n}$ when the neighbor galaxy is the early-type galaxy. Whereas
early-type fractions increase slowly with decreasing $r_{p,n}$ when 
the neighbor galaxy is a late type. This makes the early-type fraction of 
the target galaxies with the early-type neighbor larger than $0.5$ when they 
are located in the neighbor's virial radius, whereas that of the target 
galaxies with the late type neighbor becomes $\sim0.3$. This phenomenon was
interpreted as morphological conformity between close neighbors \citep{par08}. 
The gradual increase in the early-type fraction of the target galaxies with 
late-type neighbor when they come closer to the neighbor suggests that
the effect of the tidal interactions dominates other interactions such as 
hydrodynamical interactions for gas transfer from the late-type neighbor to the
target galaxy.
In the case of early-type neighbors, hydrodynamical interactions enhance
the effect of tidal interactions by removing gas to produce a rapid increase 
in the early type fraction when they become closer to the neighbor. 

%%%%%%%%%%%%%% Fig~27 > 26 > Fig26 FeTrp2vir.ps  Trp2vir.dat %%%%%%%%%%%%%%%%%%%
\begin{figure}
\plotone{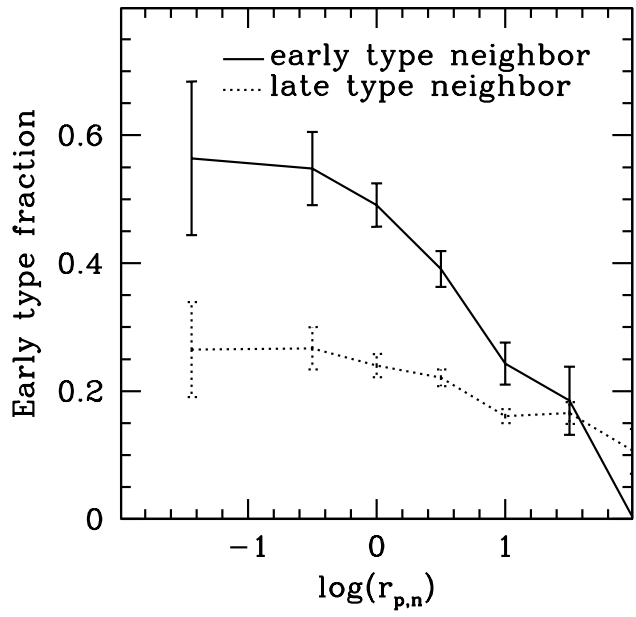}
\caption{Early-type fractions as a function of $r_{p,n}$.
\label{fig24}
}
\end{figure}
%%%%%%%%%%%%%%%%%%%%%%%%%%%%%%%%%%%%%%%%%%%%%%%%%%%%%%%%%%%%%%%%%%%%%
\clearpage

\section{Summary and Discussion}

This paper presents a catalog of the morphological types of 5836 galaxies whose
redshifts are less than $z=0.01$. The morphological types are determined by 
a visual inspection of the color images provided by SDSS DR7. The majority of
galaxies in the present sample come from the KIAS-VAGC which is
based on the spectroscopic target galaxies of the SDSS DR7 complemented
by the bright galaxies with known redshifts from various catalogs. 
Galaxies fainter than the limiting magnitude of the SDSS spectroscopic target
galaxies were also included if redshifts are available in the NED. 
The number of galaxies 
fainter than $r_{pet}=17.77$ is 809 which is approximately $14\%$ of the 
present sample. Therefore, the present 
sample is not flux-limited but a flux-limited sample can be defined if 
galaxies brighter than $r_{pet}=17.77$ are used because galaxies brighter 
than the SDSS bright limiting magnitude of $r_{pet}=14.5$ are believed to be
compiled almost completely in the NED. 

The galaxy morphology classification system adopted for bright galaxies 
in the present study is basically the de Vaucouleurs revised Hubble-Sandage
system of stages and families \citep{hub36, dev59, san61, but07}. Inner and
outer ring varieties, lenses and stages of E and S0 galaxies are not 
recognized in our catalog. For dwarf galaxies, we use notation similar
to \citet{bin85}, modified to allow for features, such as blue
cores or blue colors, that are recognizable in color images.
That is, we distinguish subtypes; dE, dE$_{bc}$, dSph, dE$_{blue}$,
and dwarf lenticulars (dS0). The dE, dSph
and dS0 galaxies are further divided into dE$_{un}$, dE$_{n}$, dSph$_{un}$,
dSph$_{n}$, dS0$_{un}$, and dS0$_{n}$ to denote the presence of nucleation. 
In the analysis of the frequency distributions, however,  we do not distinguish
them because physical properties, such as the color and luminosity are similar,
and the numbers of dE$_{un}$ and dSph$_{un}$ galaxies are much smaller than 
those of dE$_{n}$ and dSph$_{n}$ galaxies.

The present catalog contains 85 elliptical galaxies (E: 1.5\%), 1093 dwarf 
elliptical-like galaxies (dEs: 18.7\%), 286 lenticular galaxies (S0: 4.9\%), 
1874 spiral galaxies (Sp: 32.1\%), and  2498 irregular galaxies (Ir: 42.8\%).
Since most irregular galaxies are dwarf galaxies, the local universe is 
dominated by dwarf galaxies. Among giant galaxies, spiral galaxies dominate
the local universe. The late-type spirals are more frequent than the early-type
spirals and $\sim62\%$ of spiral galaxies have bars including weak bars.
A significant number of dE$_{blue}$ galaxies, 
which are generally smaller than the ordinary dE galaxies,
are included in the present catalog. Some dE$_{blue}$ galaxies have large 
cores that are somewhat bluer than the global colors. Most dE$_{blue}$ galaxies
without cores are quite compact, similar to HII region-like BCDs. The only
difference between these two types is the degree of star formation. 
HII region-like BCDs are bluer than dE$_{blue}$ galaxies.
A considerable number of dI galaxies show somewhat round shapes 
with size, luminosity, and color similar to dE$_{blue}$ galaxies. Their 
environments, represneted by the local background density, are also similar.
Therefore, we suppose that dE$_{blue}$ galaxies are the same populations as 
dI galaxies both of which served as the building blocks of giant
galaxies. 

Some evidence exists for dichotomy in the morphological properties of
dE galaxies and dSph galaxies, which
is in contrast to the conventional practice to treat them as the same class of
galaxies \citep{kor12}. The dichotomy is not as clear as that between the giant
elliptical galaxies and the dE galaxies in terms of size and 
luminosity. However, they are different in the surface brightness and color.
The colors of the dE galaxies are similar to those of giant
elliptical galaxies, whereas they differ appreciably from dSph
galaxies. The surface brightness and surface brightness gradient of dwarf 
elliptical galaxies are intermediate between the giant elliptical galaxies
and dwarf spheroidal galaxies. If we consider the kinematic difference between
dSph and dE galaxies \citep{wal09,geh10,tol12}, the origins of dSph and dE 
galaxies thought to be different, at least for the disperion-supported dS0 
galaxies and the fast rotating dE galaxies, Dispersion-supported dSph galaxies 
can be promodial objects \citep{kor14} while fast rotating dE galaxies are 
stripped late type spirals \citep{kor12}.

We explored the physical properties, such as luminosity, color, size, and 
axial ratios, of the local galaxies. In general, distributions of the
physical parameters of galaxies with different morphological types are quite
different in terms of the most probable values and the shapes of their
distributions. 
Elliptical galaxies show the highest luminosity but their maximum size is 
smaller than that of spiral galaxies.
The larger size of spiral galaxies is supposed to be due to the accretion
of the intergalactic material after the collapse phase of galaxy formation. 
The luminosity and size distributions of the two types of dwarf galaxies, 
dEs and dIr, are very similar, implying similar masses.
The dependence of galaxy size and luminosity on galaxy
morphology seems to have originated from the morphology-density relation
because the local background density determines the dynamical 
timescales ($\tau \propto \rho^{-1/2}$), which is closely related to the 
star formation rates. Elliptical
and lenticular galaxies that are likely to be formed in dense environment
are assumed to build up their bodies rapidly, due to short dynamical times,
without a late accretion phase, whereas spiral galaxies build up their outer
parts by accreting intergalactic material after the initial collapse phase.
This picture of spiral galaxy formation is in good agreement with the
inside-out growth picture of spiral galaxies. 

Although the local background density appears to be the primary parameter 
dictating the morphology of a galaxy, its role is different for giant galaxies
and dwarf galaxies. For giant galaxies, the local background density 
constrains the galaxy morphology through the star formation efficiency
in the early phase of galaxy formation. In contrast,
the local background density constrains the interaction rates
between the host galaxies and their satellites, which are mostly dwarf galaxies
because interaction rates are closely related to the orbital period that is 
determined by the dynamical timescales.
Therefore, we suppose that dEs, dSphs, and
dwarf lenticulars, which are likely to be observed in the
high-density regions, lose their gas quickly due to frequent
interactions with large galaxies. On the other hand, dE$_{blue}$ galaxies
must have fewer interactions than other dwarf galaxies to keep their gas
longer. In the case of the dE$_{bc}$, the role of local
background density is twofold. It is high enough for frequent interactions
to remove gas from the galaxy but it allows delayed accretion of gas to 
the nucleus of the dE galaxy to make the blue core. 

Most of the local galaxies have nearby companion galaxies. The mean projected
distance to the nearest neighbor $\bar{r_{p}}$ is $\sim260$kpc, which is 
similar to the virial radius of a bright spiral galaxy. Approximately $70\%$
of the local galaxies have the nearest neighbor within $r_{p}=\bar{r_{p}}$.
On the other hand, a considerable number of galaxies are extremely isolated. 
The number of extremely isolated galaxies depends on the isolation
criteria. If $r_{p}=1.5$Mpc is taken as the minimum projected separation
between a target galaxy and its nearest neighbor, the number of extremely
isolated galaxies is 113 but it becomes 66 if we consider $r_{p} > 2$Mpc 
as the criterion for the extreme isolation.

%--------------------------------------------------------------------
\acknowledgments
The authors are grateful to R.J. Buta, for valuable comments
and suggestions which greatly improved this paper.
This work was supported by the NRF Research grant 2010-0023319.

\clearpage
\end{document}